\documentclass[preprint,12pt]{elsarticle}
\usepackage[top=1in, bottom=1in, left=1in, right=1in]{geometry}

\usepackage{amssymb}
\usepackage{amsmath}
\usepackage{amsfonts}
\DeclareMathOperator*{\argmax}{arg\,max}
\usepackage{bm}

\usepackage{array}
\usepackage{tabularx}
\usepackage{booktabs} 
\usepackage{multirow}
\usepackage{enumitem}
\usepackage{balance}

\usepackage[caption=false,font=normalsize,labelfont=sf,textfont=sf]{subfig}
\usepackage{graphicx}
\usepackage{stfloats}

\usepackage{textcomp}
\usepackage{verbatim}
\usepackage{lipsum}
\usepackage{xcolor}  
\usepackage{fontawesome5}
\usepackage{orcidlink}

\usepackage{algorithm}
\usepackage{algorithmicx}
\usepackage[noend]{algpseudocode} 

\newcommand{\mbf}[1]{\mathbf{#1}}
\newcommand{\blue}[1]{\textcolor{black}{#1}}  
\definecolor{belowb1}{RGB}{245, 245, 220}  

\usepackage[utf8]{inputenc}

\journal{}
\makeatletter
\def\ps@pprintTitle{%
 \let\@oddhead\@empty
 \let\@evenhead\@empty
 \let\@oddfoot\@empty
 \let\@evenfoot\@empty
}
\makeatother

\begin{document}

\begin{frontmatter}

\title{Efficient rare event estimation for multimodal and high-dimensional system reliability via subset adaptive importance sampling} 

\author[inst1]{Sara Helal\corref{cor1}}
\author[inst1]{Victor Elvira}

\cortext[cor1]{Corresponding author\\\hspace*{1.65em}E-mail address: S.Helal@sms.ed.ac.uk}

\affiliation[inst1]{
  organization={School of Mathematics, University of Edinburgh},
  country={UK}
}
\date{} 
\begin{abstract}
Estimating rare events in complex systems is a key challenge in reliability analysis. The challenge grows in multimodal problems, where traditional methods often rely on a small set of design points and risk overlooking critical failure modes. 
Further, higher dimensions make the probability mass harder to capture and demand substantially larger sample sizes to estimate failures.
In this work, we propose a new sampling strategy, subset adaptive importance sampling (SAIS), that combines the strengths of subset simulation and adaptive multiple importance sampling. SAIS iteratively refines a set of proposal distributions using weighted samples from previous stages, efficiently exploring complex and high-dimensional failure regions. 
Leveraging recent advances in adaptive importance sampling, SAIS yields low-variance estimates using fewer samples than state-of-the-art methods and achieves pronounced improvements in both accuracy and computational cost.
Through a series of benchmark problems involving high-dimensional, nonlinear performance functions, and multimodal scenarios, we demonstrate that SAIS consistently outperforms competing methods in capturing diverse failure modes and estimating failure probabilities with high precision.
\end{abstract}


\end{frontmatter}

\section{Introduction}
\label{sec1}

In reliability engineering, rare events refer to low-probability system failures that demand careful assessment due to their catastrophic consequences \cite{ESHRA2025}.
Estimating the probability of such failures remains a critical challenge and is central to evaluating the safety and performance of complex systems.
 This probability, typically denoted as $P_f$, represents the probability that the system response will fall below a predefined acceptable threshold, thereby indicating failure. It is formulated as an integral over the domain defined by the performance function, which is influenced by a set of random variables that represent uncertainties in the system. The performance function is usually termed limit state function (LSF), and it can include one or multiple distinct failure modes \cite{HUANG2022,BREITUNG2019}. Given the nonlinear nature of the LSF and the high dimensionality of the random parameters, direct analytical evaluation of such integrals is often intractable. The challenge becomes particularly significant when multiple modes of failure exist, which is common in many practical applications, ranging from structural reliability to energy systems and aerospace engineering \cite{cadini2017estimation,morio2015estimation,wei2019structural}. These difficulties have fostered the research on alternative techniques for the effective estimation of $P_f$ and reduction of the number of simulations required. These techniques can be roughly classified into three main categories:  optimization-based approximation methods, surrogate model methods, and sampling-based methods. A comparative analysis of these methods is summarized in Table \ref{methods_comparison}.

In the first category, first- and second-order reliability methods (FORM and SORM) are the most prominent members \cite{keshtegar2017hybrid,gong2017first, lim2016post}. The estimation of the probability of failure is based on first- and second-order Taylor series expansion utilized to approximate the LSF around a reference point (also known as the most probable point or design point) \cite{zhao1999general}. The most probable point (MPP) refers to the location within the failure domain of the parameter space that has the highest likelihood of occurrence, making it the most significant contributor to the overall failure probability. However, approximation-based methods can lead to inaccurate estimates of $P_f$ for high dimensional problems or with problems with high nonlinear LSF and multiple MPPs. 

In the second category, commonly used surrogate methods include response surface method \cite{lucia2763}, polynomial chaos expansion \cite{PCK0000870,schobi2015polynomial}, support vector machine \cite{HURTADO2004271}, adaptive Kriging model \cite{zhang2022moving,bichon2008efficient}, and deep neural networks \cite{BAO2021107778}. These methods are applied to construct an approximate model (i.e., metamodel) to capture the behavior of the true  LSF, especially near failure regions where rare events occur, and replacing with a function that has less cost per run. However, surrogate model methods often require a significant number of sample points to effectively train the model, and the overall computational cost is largely influenced by the strategy used to select these sample points \cite{XIAO2020106852}. This leads to the employment of surrogate methods in conjunction with sampling methods, to correct the potential bias for very small failure probability estimation.

Sampling-based methods, such as Monte Carlo (MC) simulation \cite{Rubinstein1981SimulationAT}, are well-known probabilistic methods for reliability analysis. However, the computational complexity of MC simulation demands can become significant, particularly for very small probability levels or very rare events, requiring a huge amount of samples for accurate estimation. To overcome this drawback, numerous variance reduction techniques have been developed, including importance sampling (IS) \cite{elvira2021advances}, and its adaptive variants, adaptive importance sampling (AIS) \cite{cappe2008adaptive,martino2014adaptive} and population Monte Carlo (PMC) \cite{cappe2004population,elvira2022optimized,ELVIRA201777},  directional sampling \cite{guo2020active,grooteman2011adaptive}, line sampling \cite{PRADLWARTER2007208}, and subset simulation (SS) \cite{AU2001263}. The success of these methods relies on the choice of the so-called importance distribution (i.e., proposal) that is expected to generate samples that better explore the failure domain.  

\begin{table}[h!]
\centering
\caption{Comparison of methods for rare event estimation in reliability analysis.}
\renewcommand{\arraystretch}{1.5}	  
\resizebox{\textwidth}{!}{%
\begin{tabular}{@{}
>{\raggedright\arraybackslash}p{3.5cm}
>{\raggedright\arraybackslash}p{6.5cm}
>{\raggedright\arraybackslash}p{7.5cm}
>{\raggedright\arraybackslash}p{4.5cm}
@{}}\toprule
\textbf{} & \textbf{Advantages} & \textbf{Limitations} & \textbf{Examples} \\ \midrule

\textbf{Approximation methods} 
& \begin{itemize}[leftmargin=0pt, labelwidth=0pt, labelsep=0pt]
    \item  \hspace{3pt}Efficient for problems with well-defined failure regions.
    \item \hspace{3pt}Can provide a good approximation of the most probable failure point.
\end{itemize}   
& \begin{itemize}[leftmargin=0pt, labelwidth=0pt, labelsep=0pt]
    \item \hspace{3pt}Struggle with highly nonlinear or discontinuous limit state functions.
    \item \hspace{3pt}May fail in identifying multiple failure modes.
\end{itemize}  
& 
     First-order reliability method (FORM) \cite{gong2017first},
     Second-order reliability method (SORM) \cite{lim2016post},
     Finite element method (FEM) \cite{rashki2018low}.
\\
\midrule

\textbf{Surrogate model methods} 
& \begin{itemize}[leftmargin=0pt, labelwidth=0pt, labelsep=0pt]
    \item \hspace{3pt}Reduce computational cost by replacing expensive simulations with surrogate models.
    \item \hspace{3pt}Effective for problems where the failure domain can be captured with fewer simulations.
\end{itemize}  
& \begin{itemize}[leftmargin=0pt, labelwidth=0pt, labelsep=0pt]
    \item \hspace{3pt}May require extensive training data in high-dimensional and complex failure surfaces.
    \item \hspace{3pt}Introduce additional model uncertainty that requires careful quantification.
\end{itemize}  
& 
  Kriging metamodel \cite{PCK0000870},
   Polynomial chaos \cite{PCK0000870}, Response surface method \cite{lucia2763},
   Deep learning \cite{BAO2021107778}.
\\
\midrule

\textbf{Sampling-based methods} 
& \begin{itemize}[leftmargin=0pt, labelwidth=0pt, labelsep=0pt]
    \item \hspace{3pt}Versatile and applicable to highly nonlinear, multimodal, and discontinuous problems.
    \item \hspace{3pt}Flexible with respect to dimensionality and input distributions.
\end{itemize}  
& \begin{itemize}[leftmargin=0pt, labelwidth=0pt, labelsep=0pt]
  
    \item \hspace{3pt}Lack of diversity in sample space exploration may lead to inefficiency in multimodal failure domains.
    \item \hspace{3pt}Performance heavily depends on the quality of the proposal distribution.
\end{itemize}  
& 
    { Importance sampling \cite{elvira2021advances}, Line-sampling (LS) \cite{PRADLWARTER2007208},
    Weighted average simulation \cite{OKASHA201647},
     Cross-entropy PMC (CE-PMC) \cite{9667265Miller},
     Subset simulation \cite{AU2001263},
     Spherical subset simulation \cite{KATAFYGIOTIS2007194}.}
\\
\bottomrule
\end{tabular}
}
\label{methods_comparison}
\end{table}

Among these methods, SS has attracted much attention for computing small failure probabilities for reliability problems. Its efficiency stems from decomposing the original probability space into a sequence of nested rare event simulations (or subsets) of more frequent events in the conditional probability spaces, with the last one being the original failure event of interest. Generating conditional samples in these spaces is not a trivial task, however. Insufficient samples in each subset can result in a substantially inaccurate estimates of failure probability with high variance. Consequently, numerous enhancements in the conventional SS have been proposed in the past two decades \cite{zhao2022subset,GUO2022108762,ZHANG2024103693,ZUEV2012283,LI2015239}. One major category of SS, known as subset simulation based on importance sampling (SS-IS), is developed in \cite{SONG2009658}. The concept of IS procedure is employed to generate the conditional samples in the failure region to iteratively estimate small failure probabilities under specified levels.  However, the SS-IS method also has some limitations that motivate the present work.

 First, iterative parameter updates rely on a single proposal, and the sample with the highest target value is selected to update the next location parameter.
 This update process can cause the proposal distribution to converge to a single local failure mode, neglecting other significant modes in a multi-failure region problem. 
 Second limitation is the lack of sample diversity used to identify intermediate failure subsets, which can poorly tune the subset levels and degrade the accuracy of the estimator.
 Third, assuming a constant covariance matrix throughout iterations restricts exploration of the probability space, unlike in successful AIS methods \cite{Bugallo7974876}. Finally, the computational cost on reliability analysis is still huge in cases with more complicated performance functions, i.e., disconnected or irregular failure domains.

\vspace{5pt}
 {\noindent\textbf{Contribution and novelty.}  
In this paper, we propose a novel Monte Carlo framework, \textit{SAIS}, to evaluate system reliability in multi-failure-mode problems, a challenging scenario for most competing state-of-the-art approaches.\footnote{We published previous results in a conference version~\cite{10890184Sara}. Here, we introduce a novel recycling procedure for all past samples to improve the failure probability estimate at the last iteration. The recycle estimator can result in significant improvements, as demonstrated in~\cite{CORNUET2012}. Additionally, we propose a gradual shrinkage covariance estimator robust over high dimensions.}}
\vspace{0.5em}
\begin{itemize}

    \item We develop a hybrid framework that combines subset simulation (SS) with adaptive  importance sampling (AIS) techniques. To the best of our knowledge, no papers on reliability analysis utilizing this combination have been released to date.

    \item The adaptive mechanisms of SAIS naturally balance exploration and exploitation in the failure space across diverse scenarios unlike most competitors. This is achieved by the design choice of threshold adaptation and sample selection and reassignment strategies that accelerate convergence and dynamically refine proposals, while maintaining diversity in samples.

    \item We introduce new adaptive proposal updates and a gradual shrinkage covariance estimator to enhance performance in high dimensions and reduce weight degeneracy, even with a relatively small number of generated samples.

    \item We propose a new recycling-based estimator that reuses all past samples to improve the accuracy of the final failure probability estimate with minimal computational cost.

    \item We validate the performance of SAIS on challenging benchmark problems in reliability analysis.

\end{itemize}

The rest of the paper develops as follows. Section \ref{sec2} introduces the reliability problem and explains why this problem is challenging. Section \ref{sec3} presents the fundamental theories of adaptive and multiple importance sampling and subset simulation methods. The proposed method is provided in detail in Section \ref{NEW_METHOD}. Several numerical examples and obtained results are presented in Section \ref{sec5}.  Design choices and algorithm tuning are discussed in Section \ref{sec6}. Finally, Section \ref{conc} concludes the paper with noteworthy remarks  about the performance of the proposed method.

\section{Reliability problem}
\label{sec2}
In probabilistic reliability analysis, the behavior of the system with $d_x$-dimension input variables $\mathbf{x} \subseteq \mathbb{R}^{d_x}$ can be described by a performance function, \blue{or limit state function (LSF)}, $S(\mbf{x}): \mathbb{R}^{d_x} \rightarrow \mathbb{R}$. The performance function describes the failure event and takes positive values when the system behaves reliably and negative values when the systems fail
\[
\begin{cases} 
S(\mathbf{x}) < 0, & \text{(failure)} \\
S(\mathbf{x}) = 0, & \text{(limit state)} \\
S(\mathbf{x}) > 0. & \text{(safe)}
\end{cases}
\]
This failure criterion defines the target failure domain in the input $\mbf{x}$-space as follows:
\[
\mathcal{F}=\{\mbf{x} \in \mathbb{R}^{d_x}: S(\mbf{x})\leq 0\},
\]
which contains the set of variables $\mbf{x}$ that lead to unacceptable performance and exceed some prescribed threshold $b=0$. \blue{Let $\pi$ be the unnormalized target probability density function (pdf) of the random variable ${X}$, and $\widetilde{\pi}(\mathbf{x})=\frac{\pi(\mathbf{x})}{Z}$ is the normalized  pdf of the target under the availability of the normalizing constant $Z$.} The reliability problem be then to compute the failure probability $P_f$ expressed as
\begin{equation}
\label{TargetEQN}
    P_f \stackrel{\text{def.}}{=} \mathbb{P}(X \in \mathcal{F}) = \int \mathbb{I}_\mathcal{F}(\mbf{x})\tilde{\pi}(\mbf{x}) d\mbf{x},
\end{equation}
where $\mathbb{I}_{\mathcal{F}}$ is the indicator function, i.e.,
\[
\mathbb{I}_{\mathcal{F}}(\mbf{x})=
    \begin{cases} 
1, & \text{if } \mbf{x} \in \mathcal{F}, \\
0, & \text{if } \mbf{x} \notin \mathcal{F}.
\end{cases}
\]
\blue{The failure modes corresponding to $S(\mbf{x})$ are $\mbf{x}^*=\argmax_{\mbf{x}\in G} \widetilde \pi(\mbf{x})$, where $G = \{\mbf{x}:S(\mbf{x})=0\}$. }When the dimension $d_x$ is high (e.g., $d_x > 20$) or the failure boundary is complex (i.e., implicit or nonlinear), the required computational efforts to  construct and sample from $\tilde{\pi}(\mbf{x})$ is often large. Moreover, when the integration region $\mathcal{F}$ is composed of multiple disconnected failure modes, identifying all significant regions becomes increasingly challenging, as standard methods may fail to adapt efficiently to these multimodal scenarios. Since $\mathcal{F}$ is not explicitly known, it is impossible to evaluate the integral in Eq. (\ref{TargetEQN}) analytically. Therefore, the main goal is to devise an efficient method for the approximation of $P_f$. 

\section{Preliminaries}
\label{sec3}
In this section, methods in close relation to our development,
i.e., adaptive importance sampling \cite{Bugallo7974876,ELVIRA201777} and subset simulation \cite{AU2001263}, are briefly introduced.

\subsection{Importance sampling}

Importance sampling (IS) method is a variance reduction MC method that is easy to generalize and consists of generating \textit{weighted} samples from the target distribution $\tilde{\pi}$ for performing the desired inference \cite{luengo2020survey}. The algorithm draws $K$ samples from the importance distribution or proposal pdf $q(\mbf{x})$, $\mbf{x}_k \sim q(\mbf{x})$ for $k=1,...,K$. \blue{When $Z$ is known, the unnormalized IS estimator (UIS) can be used and is given by 
\begin{equation}
    \widehat{I}_{\text{UIS}}=\frac{1}{K} \sum_{k=1}^{K} \mathbb{I}_{S(\mathbf{x}_k) \leq 0} {w}_k,
\end{equation}
where $w_k=\frac{\tilde{\pi}(\mbf{x}_k)}{q(\mbf{x}_k)}$ are the importance weights associated with each of the i.i.d. samples. Even if $Z$ is unknown, self-normalized IS (SNIS) estimator \cite{elvira2021advances} can be used:
\begin{equation}
   \mathbb{E}_{\pi} \left[ \mathbb{I}_{S(\mathbf{x}) \leq 0} \right] \approx \widehat{I}_{\text{SNIS}} =  \sum_{k=1}^{K} \bar{w}_k \mathbb{I}_{S(\mathbf{x}_k) \leq 0},
\end{equation}
where $\bar{w}_k= \frac{w_k}{\sum_{k=1}^{K} w_k}$ are the normalized weights such that $\sum_{k=1}^{K} \bar{w}_{k}=1$.}

\subsection{Adaptive importance sampling (AIS)} \blue{The efficiency of $\widehat{I}_{\text{SNIS}}$ for a general target $\pi(\mbf{x})$ is significantly governed by the choice of proposal distribution $q(\mbf{x})$, leading to the development of adaptive importance sampling (AIS) \cite{CORNUET2012}.  AIS algorithms update a single or multiple proposals iteratively for every $t= 1, \dots, T$ iteration. In a generic setting, $K$ samples are drawn from $N$ proposals, $\{q_n^{(t)}(\mbf{x}; \boldsymbol{\mu}_n^{(t)}, \boldsymbol{\Sigma}_n^{(t)})\}_{n=1}^{N}$, each parametrized by $\boldsymbol{\mu}_n$ and $\mbf{\Sigma}_n$, which define the location and covariance matrix, respectively.
Considering a total of $NK$ samples are simulated at each iteration, AIS proceeds by appropriately weighting the samples
with $w_{n,k}^{(t)}= \widetilde{\pi}(\mbf{x}_{n,k}^{(t)})/q_{n}^{(t)}(\mbf{x}_{n,k}^{(t)})$.  
A significant advance in the weighting strategy of AIS is the introduction of deterministic mixture (DM) weighting given by
\begin{equation}
    w_{n,k}^{(t)} = \frac{\pi(\mathbf{x}_{n,k}^{(t)})}{\Psi(\mathbf{x}_{n,k}^{(t)})},
\label{DMweights}
\end{equation}
where a single equally weighted mixture proposal is denoted by \resizebox{0.33\linewidth}{!}{$\Psi(\mathbf{x}_{n,k}^{(t)}) \equiv \frac{1}{N} \sum_{n=1}^{N} q_n^{(t)}(\mathbf{x}_{n,k}^{(t)}\boldsymbol{\mu}_n^{(t)}, \boldsymbol{\Sigma}_n^{(t)})$}. The location parameters $\{\boldsymbol{\mu}_n^{(t+1)}\}_{n=1}^N$ are then iteratively adapted at each iteration $t$. 
Various approaches for adapting the family of proposal distributions have been introduced \cite{Bugallo7974876,ELVIRA201777}. A commonly used approach is the adaptation through resampling schemes, i.e., local resampling and global resampling. 
The three-step process (sampling, weighting, and resampling) is repeated until a stopping criterion is met, such as reaching a maximum number of iterations $T$ \cite{Elvira8682284}. Notably, DM weights are used in a much robust estimator, deterministic mixture population Monte Carlo (DM-PMC) \cite{ELVIRA201777}, that reads
\begin{equation}
\label{MISeqn}
    \widehat{P}_f =  \sum_{k=1}^{K} \sum_{n=1}^{N}  \bar{w}_{n,k}^{(T)} \mathbb{I}_{S(\mathbf{x}_{n,k}^{(T)}) \leq 0 } ,
\end{equation}}where $\bar{w}_{n,k}^{(T)}$ are the normalized weights at the final iteration $T$. From now on, we denote $\boldsymbol{\theta}_n = (\boldsymbol{\mu}_n, \boldsymbol{\Sigma}_n)$ to ease the notation.

\subsection{Subset simulation}
 {Subset simulation (SS) is a robust Monte Carlo (MC) simulation method designed to transform the rare event into a sequence of $T$ more frequent nested events that gradually approach the target failure domain \cite{au2007application}.} {Each intermediate level, indexed by $t=1, \dots,T$, defines a conditional event. A key component of SS is its use of a tailored Markov chain Monte Carlo (MCMC) method to generate conditional samples and estimate the intermediate conditional probabilities  $P^{(t)}$, albeit at the cost of producing dependent samples \cite{PAPAIOANNOU201666}. To this end, subset simulation-importance sampling (SS-IS) \cite{SONG2009658} introduces IS density function $q^{(t)}(\mbf{x})$ to generate the conditional samples. Details of SS-IS are referred to \cite{SONG2009658}. }

\blue{Let $\mathcal{F}=\{\mbf{x}  \in \mathbb{R}^{d_x}: S(\mbf{x})\leq 0\}$ be the target failure event, where $S(\mbf{x})$ is the performance function, and $b=0$ is the desired threshold value of failure events for a performance index in a system of interest.  The intermediate events are $\mathcal{F}^{(t)}=\{\mbf{x}: S(\mbf{x})\leq b^{(t)}\}$, $t=1, \dots, T$, where the failure thresholds are a decreasing sequence $\infty=b^{(0)} > b^{(1)} > \dots > b^{(T)} = 0$. The values of $b^{(t)}$ are chosen progressively as the $\rho$-quantile of the performance values $S(\mbf{x})$. Then the failure events satisfy the following relations:
\begin{equation}
    \mathbb{R}^{d_x} \equiv 
    \mathcal{F}^{(0)} \supset \mathcal{F}^{(1)} \supset \cdots  \supset \mathcal{F}^{(T-1)} \supset \mathcal{F}^{(T)} \equiv \mathcal{F},
\end{equation}
 and 
 \begin{equation}
     \mathcal{F}^{(T)} = \bigcap_{t=0}^{T} \mathcal{F}^{(t)},
 \end{equation}
where $\mathcal{F}^{(0)} \equiv \mathbb{R}^{d_x}$ is the initial event.
The probability of failure can be expressed as a product
of conditional probabilities}

\blue{\begin{equation}
   {P}_f  \equiv P^{(T)} \triangleq P^{(0)} \prod_{t=1}^{T} P^{(t\mid t-1)} \triangleq \prod_{t=0}^{T} P^{(t)},
\end{equation}
where $P^{(T)} \triangleq \mathbb{P}( \mathcal{F}^{(T)}  )$ is the probability of the final event at $T$, $P^{(t\mid t-1)} \triangleq \mathbb{P}(\mathcal{F}^{(t)}  \mid \mathcal{F}^{(t-1)} )$ is the conditional probability of event $\mathcal{F}^{(t)}$ given $\mathcal{F}^{(t-1)}$ for $ t=1,2, \dots, T$, and $P^{(t)} \triangleq \mathbb{P}( \mathcal{F}^{(t)})$ is the probability of failure event $\mathcal{F}^{(t)}$ estimated by 
\begin{equation}
    {P}^{(t)} = \frac{1}{K} \sum_{k=1}^{K} \mathbb{I}_{\mathcal{F}^{(t)}}\left(\mbf{x}^{(t)}_k\right) \quad t = 1,2,\ldots,T,
\end{equation}
where $K$ is the number of samples at level $t$. }

\begin{figure}[b]
    \centering
    \begin{tikzpicture}[scale=0.7]

\definecolor{pastelgreen}{RGB}{220, 255, 220}
\definecolor{pastelblue}{RGB}{200, 225, 255}
\definecolor{pastelorange}{RGB}{255, 230, 180}
\definecolor{darkgreen}{RGB}{150, 200, 150}  
\definecolor{darkorange}{RGB}{230, 150, 80} 

\begin{scope}[shift={(-11,0)}]  
    \fill[pastelgreen,fill opacity=0.5] (-2.5,-2.5) rectangle (2.5,2.5);

    \begin{scope}
        \clip (-2.5,-2.5) rectangle (2.5,2);  
        \fill[belowb1, fill opacity=0.9] plot[domain=-2.5:2.5,samples=100] (\x, {2 - 0.7*(\x)^2}) -- (2.5,-2.5) -- (-2.5,-2.5) -- cycle;
    \end{scope}
    
    \draw[thick,->] (-3.0,0) -- (3.0,0) node[below] {\small $x_2$};
    \draw[thick,->] (0,-3.0) -- (0,3.0) node[above] {\small $x_1$};
    \draw[thick, dashed, green!80!black] plot[domain=-2.5:2.5,samples=100] (\x, {2 - 0.7*(\x)^2});
    \node[scale=0.75,darkgreen!60!black] at (1.5, 2.1) {\small $S(\mbf{x}) = 0$};

    \foreach \r in {0.5, 1, 1.5, 2} {
        \draw[gray!80] plot[samples=50,domain=0:360] ({\r*cos(\x)}, {\r*sin(\x)});
    }
\end{scope}

\begin{scope}[shift={(-3.0,0)}]  
    \fill[pastelblue,fill opacity=0.5] (-2.5,-2.5) rectangle (2.5,2.5);

    \begin{scope}
        \clip (-2.5,-2.5) rectangle (2.5,0.2);  
        \fill[belowb1, fill opacity=0.9] plot[domain=-2.3:2.3,samples=100] (\x, {0.2 - 0.7*(\x)^2}) -- (2.3,-2.5) -- (-2.3,-2.5) -- cycle;
    \end{scope}
    
    \draw[thick,->] (-3.0,0) -- (3.0,0) node[below] {\small $x_2$};
    \draw[thick,->] (0,-3.0) -- (0,3.0) node[above] {\small $x_1$};
    
    \draw[thick, dashed, green!80!black] plot[domain=-2.5:2.5,samples=100] (\x, {2 - 0.7*(\x)^2});
    \node[scale=0.75,darkgreen!60!black] at (1.5, 2.1) {\small $S(\mbf{x}) = 0$};

    \draw[thick, blue] plot[domain=-1.95:1.95,samples=100] (\x, {0.2 - 0.7*(\x)^2});
    \node[blue] at (0.6, 0.6) {\small $b^{(1)}$};

    \foreach \r in {0.5, 1, 1.5, 2} {
        \draw[gray!80] plot[samples=50,domain=0:360] ({\r*cos(\x)}, {\r*sin(\x)});
    }
\end{scope}

\begin{scope}[shift={(4.8,0)}]  
    \fill[pastelorange,fill opacity=0.6] (-2.5,-2.5) rectangle (2.5,2.5);

    \begin{scope}
        \clip (-2.5,-2.5) rectangle (2.5,1.1);  
        \fill[belowb1, fill opacity=0.9] plot[domain=-2.5:2.5,samples=100] (\x, {0.8 - 0.7*(\x)^2}) -- (2.5,-2.5) -- (-2.5,-2.5) -- cycle;
    \end{scope}
    
    \draw[thick,->] (-3.0,0) -- (3.0,0) node[below] {\small $x_2$};
    \draw[thick,->] (0,-3.0) -- (0,3.0) node[above] {\small $x_1$};
    
    \draw[thick, dashed, green!80!black] plot[domain=-2.5:2.5,samples=100] (\x, {2 - 0.7*(\x)^2});
    \node[scale=0.75,darkgreen!60!black] at (1.5, 2.1) {\small $S(\mbf{x}) = 0$};

    \draw[thick, blue] plot[domain=-1.95:1.95,samples=100] (\x, {0.2 - 0.7*(\x)^2});

    \draw[thick, orange] plot[domain=-2.17:2.17,samples=100] (\x, {0.8 - 0.7*(\x)^2});
    \node[darkorange!80!black] at (0.5, 1.2) {\small $b^{(2)}$};

    \foreach \r in {0.5, 1, 1.5, 2} {
        \draw[gray!80] plot[samples=50,domain=0:360] ({\r*cos(\x)}, {\r*sin(\x)});
    }
\end{scope}

\node[darkgreen!80!black] at (-13, 2.0) {\small $\mathcal{F}$};
\node[blue] at (-4.65, 2.0) {\small $\mathcal{F}^{(1)}$};
\node[darkorange!80!black] at (3.2, 2.0) {\small $\mathcal{F}^{(2)}$};

\end{tikzpicture}
\caption{Schematic illustration of the SS algorithm. The contours represent target $\pi \sim \mathcal{N}(0, \mbf{I})$. The green dashed curve represents the LSF $S(\mbf{x})= 5-x_2-0.7(x_1-0.1)^2$. The shaded regions define the failure domains $\mathcal{F}^{(t)}=\{\mbf{x}: S(\mbf{x}) \leq b^{(t)}$\}. }
    \label{fig:SS}
\end{figure}
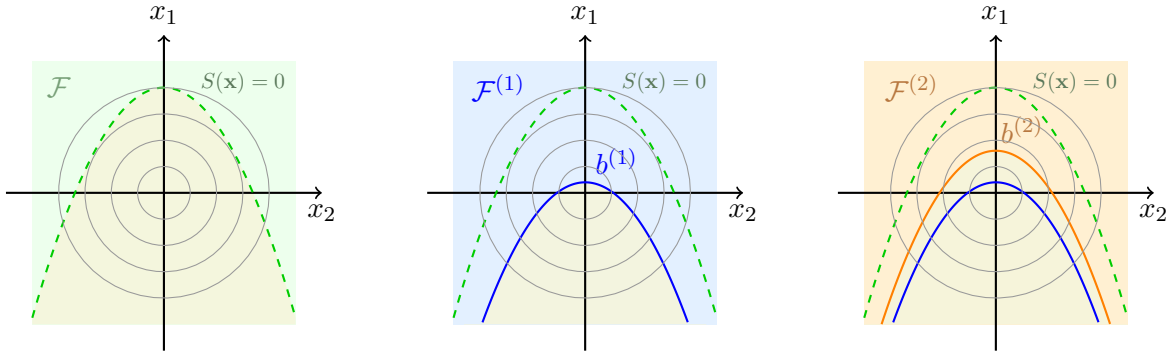

\medskip

\noindent\textbf{Toy example.}
To illustrate the practical workings of subset simulation, we now present a concrete two-dimensional toy example. Let the limit state function be defined as
$S(\mathbf{x}) = 5 - x_2 - 0.7(x_1 - 0.1)^2$, and let the reference distribution be the standard normal distribution $\tilde{\pi}(\mbf{x}) \sim \mathcal{N}(\mathbf{0}, \mathbf{I})$. Fig. \ref{fig:SS} illustrates a schematic of SS mechanism in the standard space. The green dashed curve corresponds to the true LSF $S(\mathbf{x}) = 0$, which delimits the failure region $\mathcal{F} = \{ \mathbf{x} : S(\mathbf{x}) \leq 0 \}$ shaded in green on the left panel. In the middle and right panels, the algorithm adaptively constructs a sequence of intermediate failure domains $\mathcal{F}^{(t)} = \{ \mathbf{x} : S(\mathbf{x}) \leq b^{(t)} \}$, for $t = 1$ and $2$, associated with thresholds $b^{(1)}$ and $b^{(2)}$, respectively. The target distribution $\tilde{\pi}$ is visualized through circular contours representing level sets of the standard bivariate Gaussian density. As the iteration progresses, each intermediate set $\mathcal{F}^{(t)}$ moves closer to $\mathcal{F}$ with progressively tighter level sets of $S(\mbf{x})$.

\section{The proposed SAIS algorithm}
\label{NEW_METHOD}
{We propose here the \textit{subset adaptive importance sampling} (SAIS), a failure mode search method, to accurately and efficiently solve the estimation problem in Eq. (\ref{TargetEQN}).} We present the framework in Algorithm \ref{SAISalg}.

{The algorithm proceeds in four main steps and runs through $T$ iterations. In Step 1, $K$ samples $\{\mbf{x}_k^{(t)}\}_{k=1}^{K}$ are generated at each $t$ according to a set of $N$ multivariate proposal pdfs $\{q_n^{(t)}(\mbf{x};\boldsymbol{\theta}_n^{(t)})\}_{n=1}^{N}$, with predefined means and covariances and whose sampling center lies in $\mathcal{F}^{(t-1)}$. Failure samples that lie in $\mathcal{F}^{(t-1)}$, $\mathcal{M}_n^{(t)}=\left\{ \mathbf{x}_k \mid \mathbf{x}_k \in \mathcal{F}^{(t-1)} \right\}_{k=1}^{K}$, are used as seeds to update thresholds at $t$. In Step 2, the intermediate thresholds $b^{(t)}$ are adapted to build up to the target performance threshold $b=0$. The performance function $S(\mbf{x})$ is evaluated on a permuted set of failure samples at $\mathcal{F}^{(t-1)}$. The original variable space is separated into several subsets $\mathcal{F}^{(t)}=\{\mbf{x}: S(\mbf{x}) \leq b^{(t)}\}$ $t=1,2, \dots , T$, where $\mbf{b}=(b^{(1)}, b^{(2)}, \dots, b^{(T)})$ are sorted in the descending order. In Step 3, the parameters of the importance densities are updated from $\{\boldsymbol{\theta}_n^{(t)}\}$ to $\{\boldsymbol{\theta}_n^{(t+1)}\}$ 
using the cross entropy method, and the importance weights are computed using DM weights as in Eq. (\ref{DMweights}). 
In Step 4, a new recycling scheme is incorporated to combine multiple estimates $I^{(t)}$ into the final estimator to provide a robust approximation of the sampled modes of failure.  We describe the process of each step in details in the following subsections.}    

\begin{algorithm}
\small
\caption{The SAIS algorithm }\label{SAISalg}
\resizebox{\linewidth}{!}{%
\begin{minipage}{\linewidth}
\begin{algorithmic}
\renewcommand{\algorithmicdo}{}
\renewcommand{\algorithmicwhile}{\textbf{[while}} 
\renewcommand{\algorithmicelse}{\textbf{[else}} 
\renewcommand{\algorithmicthen}{}  
\renewcommand{\algorithmicif}{\textbf{[if}} 
\State \textbf{Input:} $(N, K) \in \mathbb{N}^{+}$, $\{\boldsymbol{\mu}_{n}^{(1)}, \mbf{\Sigma}_{n}^{(1)}\}_{n=1}^{N}$,  $\mathcal{F}^{(0)} = \mathbb{R}^{d_x}$, quantile parameter $ \rho$, target threshold $b=0$, $t=1$. 

\While{$b^{(t)} \geq b$}\textbf{]}
        
 \Statex \textbf{Step 1: Sampling and seed selection.}
    \For{\textbf{\( n = 1 \text{ to } N \textbf{]} \)}}
   
    \Statex 
    \begin{enumerate}[label=(\alph*), leftmargin=3em, labelwidth=1em, align=left] 
        \item Draw $K$ samples from each proposal distribution as
        \[
        \{\mbf{x}_{n,k}^{(t)}\}_{k=1}^{K} \sim q_{n}^{(t)}(\cdot; \boldsymbol{\mu}_n^{(t)}, \boldsymbol{\Sigma}_n^{(t)}).
        \]
        \item Identify failure samples \( \mathcal{M}_n^{(t)} = \{ \mbf{x}_{n,k}^{(t)} \in \mathcal{F}^{(t-1)} \} \) and set \( M^{(t)} = |\mathcal{M}_n^{(t)}| \leq K \).
        \item Find elites \( \mathcal{A}_n^{(t)} = \{ \mbf{x}_{n,m}^{(t)} \}_{m=1}^{\lfloor\rho M^{(t)}\rfloor} \) corresponding to the top $\rho M$ samples in Step 1c.
    \end{enumerate}
        \Statex
    \EndFor
    \Statex \textbf{Step 2: Threshold adaptation.}
     \Statex 
    \begin{enumerate}[label=(\alph*), leftmargin=2em, labelwidth=1em, align=left] 
        \item Combine elites \( \mathcal{A}^{(t)} = \bigcup_{n=1}^{N} \mathcal{A}_n^{(t)} \), \( A^{(t)} = |\mathcal{A}^{(t)}| \).
        \item Define the ordered set $\{\widetilde{\mbf{x}}^{(t)}_k\}_{k=1}^{A^{(t)}}$ as a permutation of $\{\mbf{x}^{(t)}_k\}_{k=1}^{A^{(t)}}$ such that $S(\widetilde{\mbf{x}}_{1}^{(t)}) \geq S(\tilde{\mbf{x}}_{2}^{(t)}) \geq \ldots \geq S(\widetilde{\mbf{x}}_{A^{(t)}}^{(t)})$. 
        \item Let threshold \( b^{(t)}= S(\widetilde{\mbf{x}}_{\lfloor \rho A^{(t)} \rfloor}^{(t)}) \) be the $\rho$ sample quantile of the performances and \( \mathcal{F}^{(t)} = \{ \mbf{x} : S(x) \leq b^{(t)}\} \).
    \end{enumerate}

    \Statex
    \Statex \textbf{Step 3: Proposal adaptation.}
     \Statex 
    \begin{enumerate}[label=(\alph*), leftmargin=2em, labelwidth=1em, align=left] 
        \item Compute the posterior $\delta_n$ in Eq. (\ref{posterEQ}) for every $n$-th proposal using $\{\mbf{x}^{(t)}_k\}_{k=1}^{K}$ and reassign samples $K^*$ according to ${\arg\max} \ \delta_n(\mbf{x}_{k}^{(t)})$.
        \item Use the new sample set and determine the DM-weights $\{w_{n,k}^{(t)} \}_{n,k=1}^{N,K^*}$ with respect to Eq. (\ref{DMweights}).
         \item \textbf{[if} \(\text{ESS} \geq N_{\text{T}}\)\textbf{]}
        \item[] \hspace{1em} \parbox[t]{\dimexpr\linewidth-\algorithmicindent}{Solve the cross-entropy update using Eq. (\ref{muUpdate}), (\ref{SigmaUpdate}), and (\ref{Shrink_cov}) and normalized weights to obtain $\{\boldsymbol{\mu}_n^{(t+1)}, \mbf{\Sigma}_n^{(t+1)})\}_{n=1}^{N}$.}
    
        \item \textbf{[else]} 
        \item[] \hspace{1em} \parbox[t]{\dimexpr\linewidth-\algorithmicindent}{Transform the weights $w_{n,k}^{(t)*}$ using Eq. (\ref{transW}) and compute the tempered mean \(\boldsymbol{\mu}_{n}^{(t+1)*}\) using Eq. (\ref{muUpdate}) and the covariance matrix \(\boldsymbol{\Sigma}_{n}^{(t+1)*}\) using Eq. (\ref{NEWcov_t}) and (\ref{Shrink_cov}).}
    \end{enumerate}
    \EndWhile
    \Statex \textbf{Step 4: Failure estimation.}
    \Statex 
    \begin{enumerate}[label=(\alph*), leftmargin=2em, labelwidth=1em, align=left] 
    \item {Compute the intermediate failure probabilities}
    \[
 \widehat{I}_f^{(t)} = \frac{1}{NK} \sum_{k=1}^{K} \sum_{n=1}^{N}  w_{n,k}^{(t)} \ \mathbb{I}_{S(\mathbf{x}_{n,k}^{(t)}) \leq 0 },
\]
and set $\alpha^{(t)}= \lambda^{(T-t)}$; $\lambda \in (0,1)$.
\item Set $t = t + 1$.
\end{enumerate}
\Statex \textbf{[end while]}
\State \textbf{Output:} 
\[
\widehat{P}_f= \frac{1-\lambda}{1-\lambda^{(T)}}\sum_{t=1}^{T} \alpha^{(t)} \widehat{I}_f^{(t)}.
\]
\end{algorithmic}
\end{minipage}
}
\label{AlgorithmNEW}
\end{algorithm}
\subsection{Threshold adaptation (Step 2)}
\label{btADAPT}
We now describe a threshold adaptation strategy that controls the position of the sequence $b^{(t)}$. 
Let $\mathcal{X}^{(t)}$ be the total set of samples at iteration $t$, and $\mathcal{X}^{(t)}_n= \{\mbf{x}_{n,k}^{(t)}\}_{k=1}^{K}$ be the subset of samples generated by each proposal $q_n$, i.e., 
\begin{equation*}
    \mathcal{X}^{(t)} = \bigcup_{n=1}^{N} \mathcal{X}^{(t)}_n, \quad \mathcal{X}^{(t)}_n \cap \mathcal{X}^{(t)}_m = \emptyset \quad \text{for } n \neq m.
\end{equation*}

\begin{figure}[h]
	\centering
	\includegraphics[width=0.65\columnwidth]{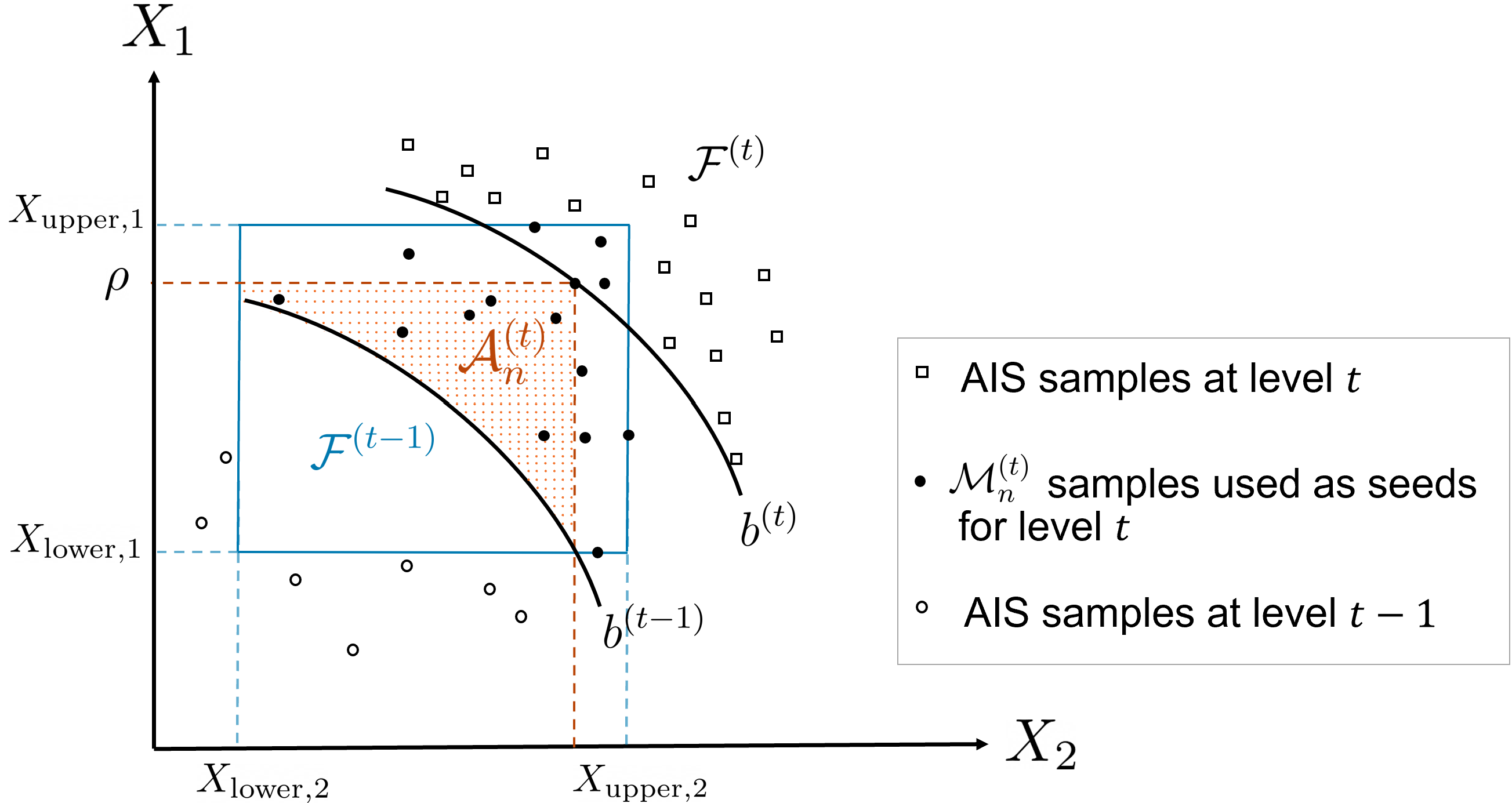}
	\caption{Sample selection representation for threshold adaptation in two subsequent SS levels. $\left (X_{\text{lower}}, X_{\text{upper}}\right) \quad$define the limits of the failure samples in subset $\mathcal{F}^{(t-1)}$.}
	\label{FIG_SEEDS}
\end{figure}

Define the initial level, $\mathcal{F}^{(0)} \equiv \mathbb{R}^{d_x}$, and $\mathcal{M}_n^{(t)} = \{ \mathbf{x}^{(t)}_{n,k} \in \mathcal{F}^{(t-1)} \} \subseteq \mathcal{X}^{(t)}_n$ as the set of failure samples within each $\mathcal{X}^{(t)}_n$ in $\mathcal{F}^{(t-1)}$, and $M^{(t)}=|\mathcal{M}_n^{(t)}|$. We rank $\mathcal{M}_n^{(t)}$ according to the values of $S(\cdot)$ and select the top $\mathcal{A}_n^{(t)} = \{ \mathbf{x}^{(t)}_{n,m} \}_{m=1}^{\lfloor \rho M^{(t)}\rfloor}$ samples (`elites'). The performance function is then evaluated at a permuted set $\widetilde{\mathcal{A}}_n^{(t)} = \{ \widetilde{\mathbf{x}}_k^{(t)} \}_{k=1}^{A^{(t)}}$, such that
\[
S(\widetilde{\mathbf{x}}_1^{(t)}) \geq S(\widetilde{\mathbf{x}}_2^{(t)}) \geq \cdots \geq S(\widetilde{\mathbf{x}}_{A^{(t)}}^{(t)}),
\]
where $A^{(t)} = |\bigcup_{n=1}^{N} \mathcal{A}_n^{(t)}|$ is the total number of elites across all proposals.
Subsequently, we update $b^{(t)}$ as the $\lfloor \rho A^{(t)} \rfloor$th largest integer value (i.e., an order statistic) of $S(\widetilde{\mbf{x}}_{k}^{(t)})$, $k=1, \dots, A^{(t)}$, yielding
 \[
 b^{(t)}= S(\widetilde{x}_{\lfloor \rho A^{(t)} \rfloor}^{(t)}),
 \]
  with the updated failure domain $\mathcal{F}^{(t)}=\{\mbf{x}: S(\mbf{x}) \leq b^{(t)}\}$. 
  Fig. \ref{FIG_SEEDS} illustrates the selection of elite samples $\mathcal{A}_n^{(t)}$ and failure samples $\mathcal{M}_n^{(t)}$ as seeds for the next level. Having introduced the first step in SAIS, we now proceed to formalize a robust proposal adaptation framework.

\subsection{Proposal adaptation (Step 3)}
\label{ProposalAdapt}
In this paper, we employ the cross-entropy method to adaptively refine a sequence of IS densities $\{q_n^{(t)}(\mbf{x})\}_{n=1}^{N}$ for $t=1, \dots, T$. Each step progressively reduces the estimator error.
\vspace{5pt}

\noindent\textbf{Sample reassignment.} We introduce a deterministic local search step that explores the sample neighborhood of each $n$-th proposal and reallocates failure samples $\mathcal{M}_n^{(t)}=\{ \mbf{x}_{n,k}^{(t)} \in \mathcal{F}^{(t-1)}\}$ at iteration $t$ based on hard reassignment weights.
For each data point $\mbf{x}_{n,k}^{(t)} \in \mathcal{M}_n^{(t)}$: (i) compute the posterior probability $\delta_n$ using Eq. (\ref{posterEQ}), and (ii) assign a binary proposal indicator $\xi_{n,k}^{(t)} \in \{0,1\}$ to denote that the $k$-th sample belongs to the $n$-th proposal with the highest posterior $\delta_n$.

The posterior probability $\delta_n(\mbf{x}_{n,k}^{(t)};  \boldsymbol{\theta}_n^{(t)})$ for the $n$-th proposal is given by 
\begin{equation}
\label{posterEQ}
\delta_n(\mbf{x}_{n,k}^{(t)};  \boldsymbol{\theta}_n^{(t)}) = \frac{q_n^{(t)}(\mbf{x}_{n,k}^{(t)}; \boldsymbol{\theta}_{n}^{(t)})}{\sum_{j=1}^{N}  q_j^{(t)}(\mbf{x}_{j,k}^{(t)}; \boldsymbol{\theta}_{j}^{(t)})},
\end{equation}
where $q_n^{(t)}(\mbf{x}_{n,k}^{(t)}; \boldsymbol{\theta}_{n}^{(t)})$ is the likelihood of the $k$-th sample under the $n$-th proposal for $k= 1, \ldots, NK $ and $n = 1, \ldots, N$.  The binary proposal indicator is then defined by
 \begin{equation}
{\xi_{n,k}^{(t)}}=
\begin{cases}
1 & \text{if } n = \underset{1 \leq n \leq N}{\arg\max} \ \delta_n(\mbf{x}_{n,k}^{(t)};  \boldsymbol{\theta}_n^{(t)}), \\
0 & \text{otherwise}.
\end{cases}
\end{equation}

\noindent\textbf{Parameter updates.}
We now outline the construction and update of intermediate mixture densities using reassigned failure samples $K^*$ from the previous step.
Specifically, from iteration $t$ to $t+1$, failure samples \( \mathcal{M}_n^{(t)}\) are used to update the parameters  $\boldsymbol{\theta}_n^{(t)} = (\boldsymbol{\mu}_n^{(t)}, \boldsymbol{\Sigma}_n^{(t)})$ to minimize the Kullback-Leibler
divergence between the optimal sampling density and each proposal
$q_n^{(t)}$. This adaptation can be employed in a cross-entropy fashion which results in $n=1,2, \dots, N$ optimization problems, each given by
\begin{equation}
\max_{\boldsymbol{\theta}_n} \frac{1}{K} \sum_{k=1}^{K} w_{n,k}^{(t)} \ \mathbb{I}_{S(\mathbf{x}_{n,k}^{(t)}) \leq {b}^{(t)}} \ \log \left( q_n^{(t)}(\mathbf{x}^{(t)}_{n,k}; \boldsymbol{\theta}_n^{(t)}) \right),
\label{CEopt}
\end{equation}
where $w_{n,k}^{(t)}$ are the deterministic mixture (DM) weights \eqref{MISeqn}.

Following the solution to the update Eq. (\ref{CEopt}), the mean and covariance parameters of every $n$-th proposal for the next iteration $t+1$ are given as \cite{10890184Sara} 
\begin{align}
    \boldsymbol{\widehat{\mu}}_n^{(t+1)} &= 
    \frac{\sum_{k=1}^{K} w_{n,k}^{(t)} \mathbb{I}_{S(\mathbf{x}_{n,k}^{(t)}) \leq {b}^{(t)}}  \xi_{n,k}^{(t)} \mathbf{x}_{n,k}^{(t)}}{\sum_{k=1}^{K} w_{n,k}^{(t)} \mathbb{I}_{S(\mathbf{x}_{n,k}^{(t)}) \leq {b}^{(t)}}  \xi_{n,k}^{(t)}}, \label{muUpdate} \\
    \mathbf{\widehat{\Sigma}}_n^{(t+1)} &=
\textstyle{\frac{\sum_{k=1}^{K} w_{n,k}^{(t)}  \mathbb{I}_{S(\mathbf{x}_{n,k}^{(t)}) \leq {b}^{(t)}} \xi_{n,k}^{(t)} \left(\mathbf{x}_{n,k}^{(t)} - \boldsymbol{\mu}_n^{(t)} \right) \left(\mathbf{x}_{n,k}^{(t)} - \boldsymbol{\mu}_n^{(t)} \right)^{\top}}{\sum_{k=1}^{K} w_{n,k}^{(t)} \mathbb{I}_{S(\mathbf{x}_{n,k}^{(t)}) \leq {b}^{(t)}} \xi_{n,k}^{(t)}},} \label{SigmaUpdate}
\end{align}
respectively.

\noindent\textbf{Covariance learning strategy.}
Our criterion for robust covariance learning against weight degeneracy rests upon ensuring that the effective sample size  ${\text{ESS}} = ( \sum_{k=1}^K ( \bar{w}_{n,k}^{(t)} )^2)^{-1}$ at each iteration $t$ meets a specified lower threshold $N_{\text{T}}=K^*/2$, where $K^*$ is the number of samples after reassignment \cite{delmoral41462}. If the local ESS $ > N_{\text{T}}$ at $t$, we update the empirical covariance using Eq. (\ref{SigmaUpdate}) and normalized DM weights. Otherwise, the unnormalized weights $w_{n,k}^{(t)}$ are transformed to 
    \begin{equation}
    \label{transW}
   w_{n,k}^{(t)*} = \psi(w_{n,k}^{(t)})=\left(w_{n,k}^{(t)}\right)^{{\gamma_t}},
    \end{equation}
   for $k = 1,\ldots,K^*$ and $n=1, \ldots, N$, where $\psi$ is the tempering transformation function defined by an increasing sequence $\gamma_t $, i.e., $0\leq \gamma_t \leq 1$ \cite{ElLaham2018RobustCA}. 
   We construct $\gamma_t$ using sigmoid function  $\gamma_t = \frac{1}{1+e^{-t}}$, depending on the iteration index $t$. 
   The normalized transformed weights $\bar{w}_{n,k}^{(t)*}$ are then used for the mean adaptation $\widehat{\boldsymbol{\mu}}_{n}^{(t)*}$ in Eq. (\ref{muUpdate}), and the covariance is updated as
    \begin{equation}    \resizebox{0.65\textwidth}{!}{${\boldsymbol{\widehat{\Sigma}}}_n^{(t+1)*} = \frac{\sum_{k=1}^{K} \bar{w}_{n,k}^{(t)*} \mathbb{I}_{S(\mathbf{x}_{n,k}^{(t)}) \leq {b}^{(t)}}  \xi_{n,k}^{(t)} \left(\mathbf{x}_{n,k}^{(t)} - \boldsymbol{\mu}_{n}^{(t)*} \right) \left(\mathbf{x}_{n,k}^{(t)} - \boldsymbol{\mu}_{n}^{(t)*}\right)^{\top}}{\sum_{k=1}^{K} \bar{w}_{n,k}^{(t)*} \mathbb{I}_{S(\mbf{x}_{n,k}^{(t)}) \leq {b}^{(t)}}  \xi_{n,k}^{(t)}}.$}
    \label{NEWcov_t}
    \end{equation}

Following \cite{9022450LAHAM}, we propose to further stabilize the covariance updates at higher dimensions $d_x$ and apply a
covariance shrinkage approach 
\begin{equation}
\label{Shrink_cov}
    \mbf{\Sigma}_n^{(t+1)} = (1 - \beta^{(t)}) \mbf{\Sigma}_n^{(t)} + \beta^{(t)} \widehat{\mbf{\Sigma}}_n^{(t+1)} + \eta^{(t)} \widetilde{\mbf{\Sigma}}_n^{(t+1)},
\end{equation}
where $\widehat{\mbf{\Sigma}}_n^{(t+1)}$ is the intermediate weighted covariance defined in Eqs. (\ref{SigmaUpdate}) and (\ref{NEWcov_t}) based on $N_{\text{T}}$ \cite{delmoral41462}, $\widetilde{\mbf{\Sigma}}_n^{(t+1)} = \frac{\text{tr}(\widehat{\mbf{\Sigma}}^{(t+1)})}{d_x} \mbf{I}_{d_x}$  is an isotropic diagonal empirical covariance matrix, and $\mbf{I}_{d_x}$ is the identity matrix of dimension $d_x$, $0<\beta^{(t)} \leq 1$ is the Ledoit Wolf (LW) shrinkage coefficient, and $\eta^{(t)}= 0.1t^{-1}$  is a decreasing sequence of constants controlled by $t$. 
Denoting $\widehat{\mbf{\Sigma}}^{(t+1)}_n$ by $\widehat{\mbf{S}}_n$, the optimal LW shrinkage coefficient is given by \cite{chen2010shrinkage}
\begin{equation}
    {\beta}^{(t)} =
\frac{\sum\limits_{k=1}^{K} \left\| \mathbf{x}_{n,k} \mathbf{x}_{n,k}^\top - \widehat{\mathbf{S}}_n \right\|_F^2}
{K^2 \left( \text{tr}(\widehat{\mathbf{S}}_n^2) - \frac{\text{tr}^2(\widehat{\mathbf{S}}_n)}{d_x} \right) }.
\end{equation}

\subsection{Failure probability estimation (Step 4)}

\noindent\textbf{Recycled AIS.} 
To mitigate residual uncertainty in the last threshold $b^{(T)}$, we propose recycling all samples generated across $T$ iterations of the SAIS process \cite{nguyen2014improving}. 
 Here we assume each iteration provides an approximation $\widehat{I}_f^{(t)}$ of
the target (\ref{TargetEQN}) with a weighted empirical distribution. The weighting factor, also known as forgetting factor, is given by
\begin{equation}
    \alpha^{(t)}= \lambda^{(T-t)}, \quad \quad t=1,2,\dots, T,
\label{alpha}
\end{equation}
assigns exponentially decreasing weights to earlier iterations, giving greater influence to recent samples in the estimator.
$\lambda \in (0,1)$ is a positive constant close to, but greater than 0, i.e., lower $\lambda$ values weight recent estimations more heavily. Note that the weights in Eq. (\ref{alpha}) do not sum to $1$ for a finite $T$. To remedy this, we modify the estimator (\ref{MISeqn}) combining multiple estimates $\widehat{I}_f^{(t)}$ into a single one as
\begin{equation}
    \widehat{P}_f= {\alpha}\sum_{t=1}^{T} \alpha^{(t)} \widehat{I}_f^{(t)},
\label{NEWestimator}
\end{equation}
with the normalization constant
\begin{equation}
    \alpha=\left (\sum_{\tau=1}^{T}\lambda^{(T-\tau)}\right )^{-1}=\frac{1-\lambda}{1-\lambda^{(T)}}.
\end{equation}

\section{Design choices and discussion}
\label{sec5}

\subsection{Tuning the algorithm}
We now provide general guidelines and remarks for setting algorithm parameters.
\begin{enumerate}[label=(\alph*), leftmargin=*, align=left]
\item \textit{Choice of $b^{(t)}$:} If the difference $b^{(t)}-b^{(t-1)}$ is too large, the sequence will converge too rapidly and will require more samples $K$ to obtain an accurate estimate of the $P_f$ in each iteration, which again increases the total number of samples. If, on the contrary, the consecutive values are chosen too close, the sequence will converge slowly and the algorithm will take a large total number of iterations $T$, i.e., the computational effort is high to progress to the target failure regions of  interest. A choice that achieves a good tradeoff is to `adaptively' choose the $\mbf{b}$-sequence, i.e., the $\rho$-quantile of the performance.
 \item \textit{Choice of $\alpha^{(t)}$:} The weight coefficients \(\alpha^{(t)}\) are selected to minimize the variance of the recycled estimator (\ref{NEWestimator}) relative to the non-recycled version (\ref{MISeqn}). Several approaches have been proposed in the literature \cite{Owen2019TheSR,DoucMINvar2007,6884589LE}, but many require extensive recalibrations and exhibit limitations in recycled importance sampling. We propose to use the exponential weighting $\alpha^{(t)} = \lambda^{(T-t)}$, which involves a single calibration phase, enhancing the overall computational efficiency.

 \item \textit{Choice of $\rho$:} Note that one needs to choose an appropriate quantile parameter $\rho$. If it is too small, the computational cost will be too high because the number of intermediate levels $b^{(t)}$ should increase accordingly to ensure that each mixture component is sampled; if too large, the mixture density may not cover the significant failure regions of the target. It is suggested in \cite{AU2001263} to use $\rho \in [0.1,0.3]$, which has been found to yield good efficiency. Therefore, the choice of $\rho$ value only has an effect on the efficiency of the method and does not influence the accuracy of the estimation.
\item \textit{Choice of $\eta^{(t)}$:} The role of $\eta^{(t)}$ is to regulate the contribution of the diagonal term $\widetilde{\mbf{\Sigma}}_n^{(t+1)}$ and ensure the updated covariance estimate remains numerically stable and well-conditioned. Two formulations can be considered $\eta^{(t)}= 0.1t^{-1}$ and $\eta^{(t)}= \beta^{(t)}t^{-1}$. Both produce strong regularization when empirical covariance is unreliable, with the former vanishing asymptotically $\lim_{t \to \infty} 0.1 t^{-1} = 0,$ and the latter keeping small but nonzero regularization even at late $t$.
 \item \textit{Choice of weight degeneracy measures:} While the ESS has some drawbacks \cite{Elvira12500}, it allows us to evaluate the accuracy of the estimator. When the ESS falls below a predefined threshold $N_{\text{T}}$, our algorithm triggers tempered updates to both the covariance $\boldsymbol{\Sigma}$ and mean $\boldsymbol{\mu}$, ensuring robustness in the adaptation process. Alternative ESS approximations, such as $\max(\bar{w}_{k}^{(t)})^{-1}$, are also possible \cite{MARTINO2017386}. In our experiments, we rely on  $ \text{ESS}$ as the primary measure.
 \end{enumerate}

 \subsection{Discussion}
 It has been shown
that deterministic mixture (DM) weighting in place of the standard importance sampling weights achieves better results. Heuristically, the denominator in the DM weights promotes diversity between proposal distributions \cite{ELVIRA201777}. Samples generated by a particular proposal receive higher weights if they are more distant from those produced by other proposals \cite{9667265Miller}. This approach effectively ensures greater separation between the proposals and reducing overlap by emphasizing contributions from distinct regions of the sample space.

Another important consideration is a key distinction between the standard subset simulation (SS) method and our approach, which lies in candidate selection for threshold updates. In SS, the threshold $b^{(t)}$ is set using all failure samples in $\mathcal{F}^{(t-1)}$, which can introduce bias by favoring extreme values of $S(\mbf{x})$. Furthermore, if a large fraction of samples in $\mathcal{F}^{(t-1)}$ originates from a single proposal, the method suffers from loss of diversity, causing one proposal to dominate the threshold adaptation process.
Our approach mitigates this issue by selecting permuted failure samples within the quantile $\rho$ of the performance distribution and grouping them by spatial locality. This improves sample propagation, ensures a balanced contribution from multiple proposals, and enhances threshold estimation.

At the implementation level, when the number of samples $K^*$ reassigned to $q_n^{(t)}$ is less than the effective sample size ESS, the covariance estimator (\ref{SigmaUpdate}) becomes inconsistent and suffers from weight degeneracy, indicated by \({\text{ESS}} \ll K^*\).  The combination of estimators in Eq. \eqref{Shrink_cov} and the introduction of the lower-variance estimator $\widetilde{\mbf{\Sigma}}_n^{(t+1)}$ have thus been shown to lead more stability in the incremental covariance estimate approach, governed by $t$.
In particular, the term $\widetilde{\mbf{\Sigma}}_n^{(t+1)}$ in Eq. \eqref{Shrink_cov} ensures the covariance estimate is always well conditioned, i.e,  positive definite ${\boldsymbol{\Sigma}}_n^{(t+1)} \succ 0$ and nonsingular, $\det\left({\boldsymbol{\Sigma}}_n^{(t+1)}\right) \neq 0$. As a result, the iterative process can proceed without requiring restarts, making the proposed iteration suitable for high-dimensional estimation problems.

\section{Numerical examples}
\label{sec6}
In this section, the performance of the proposed SAIS method is demonstrated by four challenging numerical examples without closed form solution. The focus here is on the complex geometric structure of the failure domains and the associated design points. In particular, we tackle, in the first example, a two-dimensional visual inspection of how the proposals evolve to iteratively capture failure domains. The second example illustrates the computational behaviour of SAIS in estimating failure probabilities in systems when failure modes exceed the number of most probable points. The third example evaluates the ability of SAIS to handle complex, disconnected failure domains with relatively high failure probability. Lastly, the fourth example demonstrates SAIS performance in high-dimensional settings.

For all examined examples, a reliable ground truth was obtained via direct MC simulation with a very high number of samples $(10^{10})$ in order to compare the results obtained by the proposed method. All the performance functions in the examples are described in the uncorrelated space proportional to $ \mathbb{I}_{S(\mathbf{x}) \leq 0} \pi(\mathbf{x}),$
where the target $ \pi(\mathbf{x}) $ is a standard multivariate Gaussian distribution. 
To apply the proposed method to non-standard Gaussian random variables, a proper transformation \cite{Michael2777,Armen1986,rackwitz1978structural} can be employed.  The iterative process of estimating $P_f$ is stopped when the intermediate threshold $b^{(t)}\leq 0$. For all examples, the SAIS algorithm is run with $\rho=0.1$.  For $t = 1, \dots, T$,
 the parameters of the covariance adaptation are set according to Section \ref{ProposalAdapt}.
 The efficiency of SAIS is examined through comparison with the classical SS-IS and CE-PMC methods.  
We compute the relative root mean square error (RRMSE) in the estimation of $P_f$ for the first three examples and the results are averaged
over 100 MC simulations. Example 4 is evaluated by the relative mean absolute logarithmic error (MALE) to account for underestimation and overestimation penalties.

\subsection{Example 1: Two-dimensional example with three failure regions}
\begin{figure*}[t]
    \centering
    \includegraphics[width=0.333\linewidth]{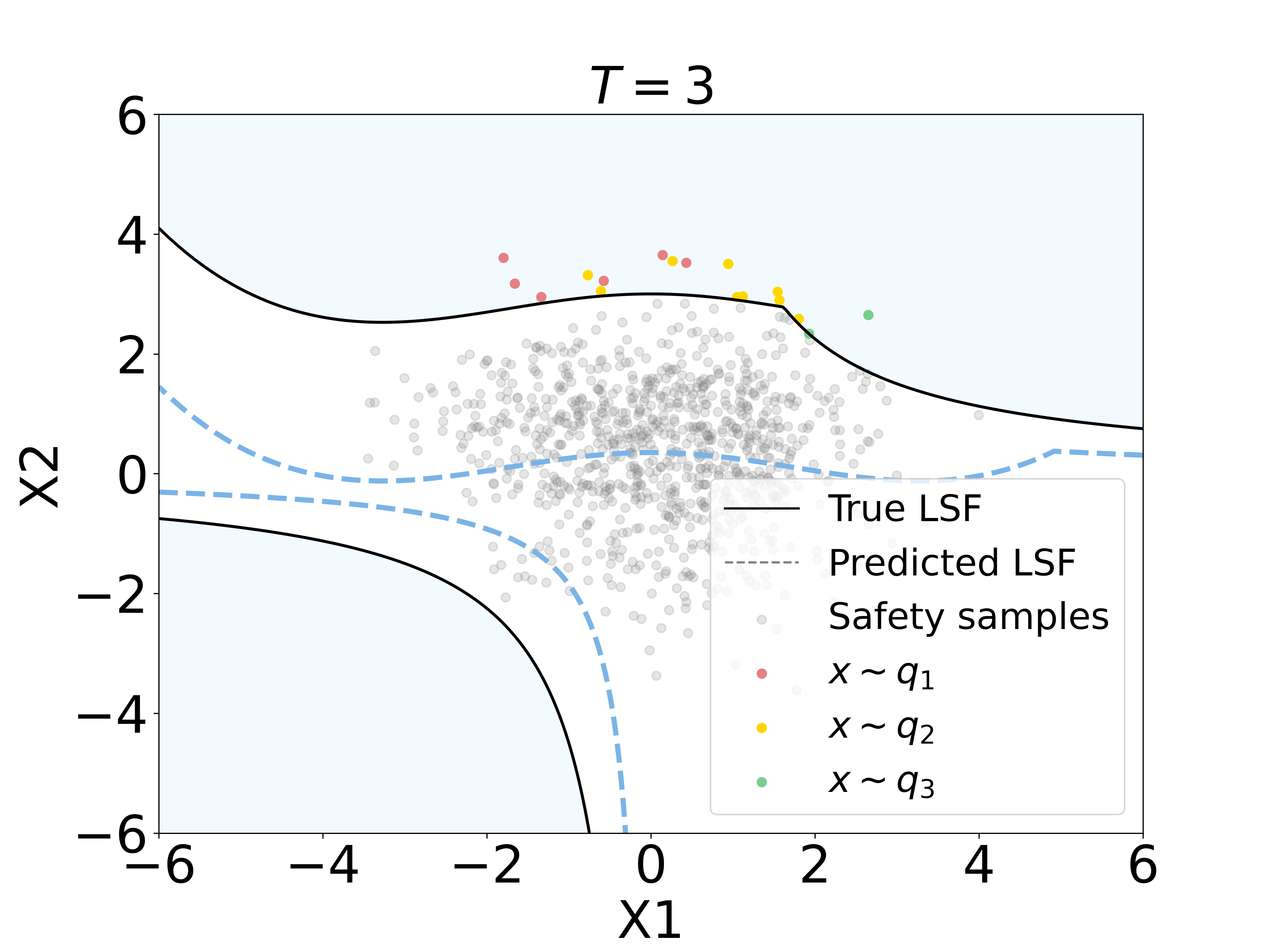}%
    \includegraphics[width=0.333\linewidth]{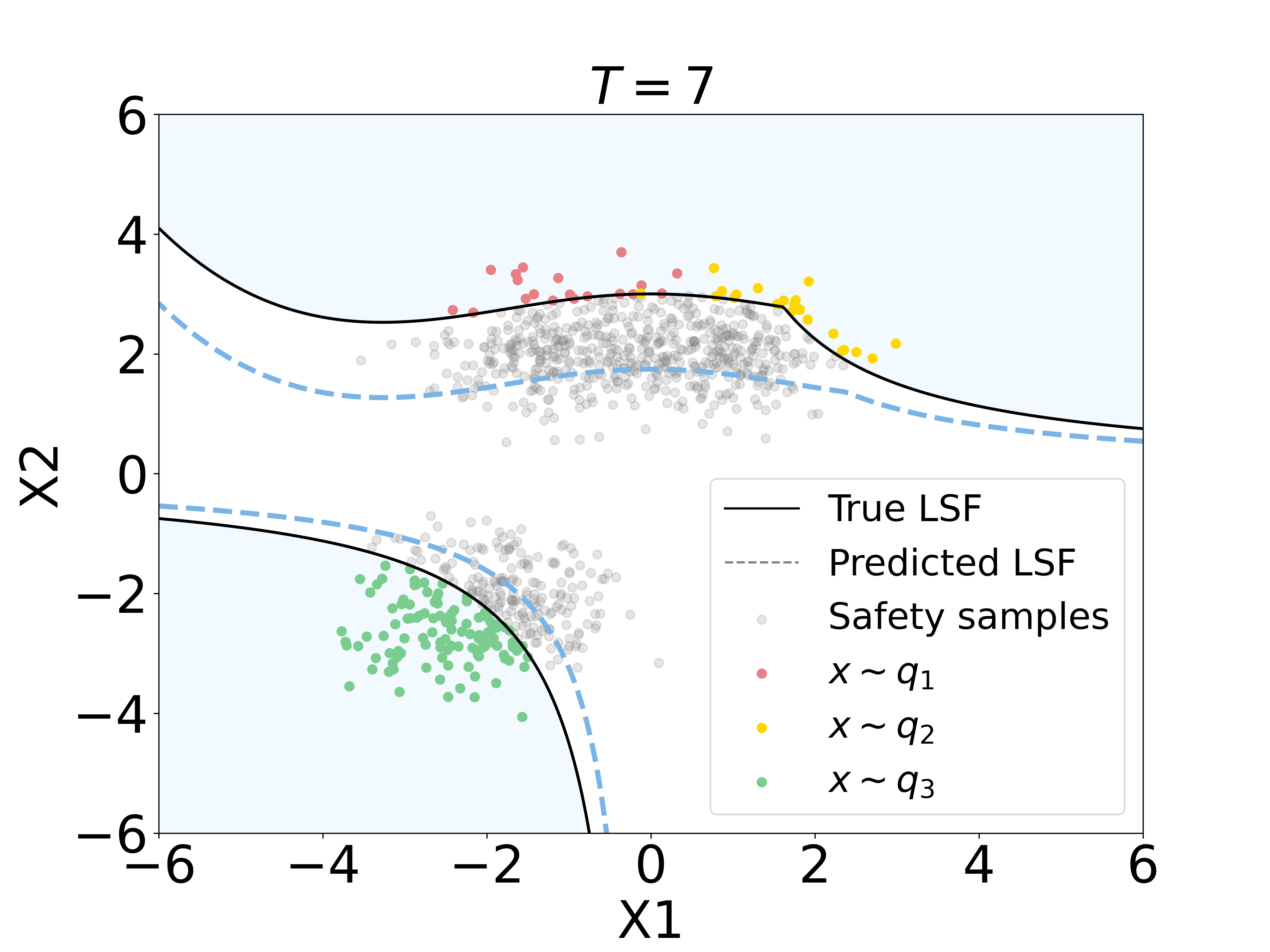}%
    \includegraphics[width=0.333\linewidth]{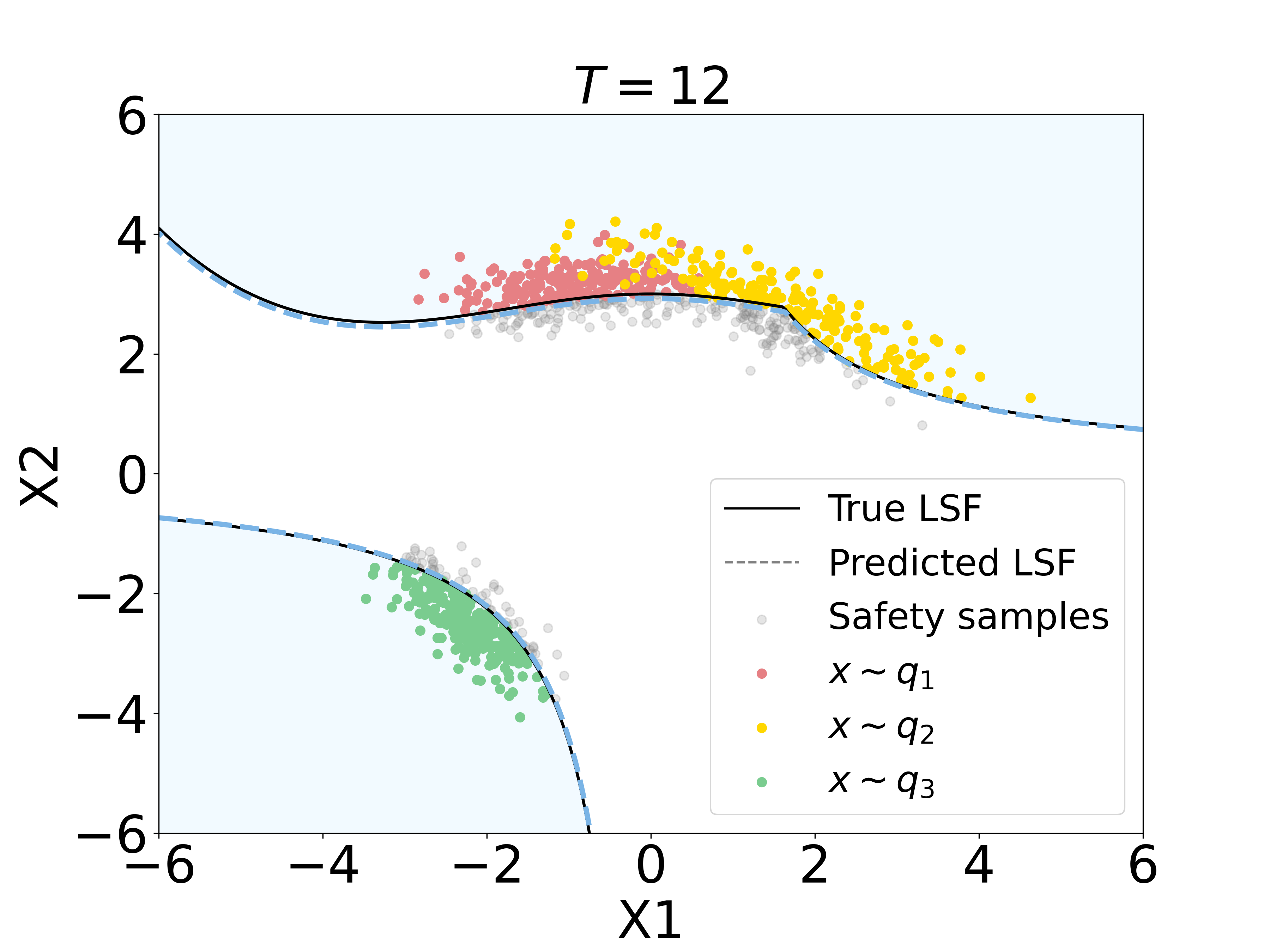}
    
    \caption{\textbf{Example 1.} The progression of $N=3$ proposals using SAIS. The plots compare the true LSF $S_1(\mbf{x})$ (solid black line) and the predicted LSF $b^{(t)}$ (dashed blue line) at three iterations, $T=3, 7,$ and $12$. The shaded regions indicate failure domains $\mathcal{F}$ where $S_1(\mbf{x})\leq 0$. The gray points indicate samples in the safe domain $S_1(\mbf{x})> 0$. Colored points represent samples generated from the mixture of proposals at $t$.}
    \label{fig_EXAMPLE1}
\end{figure*}

In this example, taken from \cite{yang2020}, we  investigate a well-known challenge in AIS: the successful evolution (i.e., survival) of proposals as the algorithm progresses \cite{ELVIRA201777}. That is, the attention is focused on whether the proposals are able to populate the failure regions and thus lead to a good estimation of the failure probability. We consider a two-dimensional case (i.e., $d_x=2$) to permit an insightful illustration of the algorithmic flow of SAIS. The performance function is defined as 
\begin{equation}
S_1(\mathbf{x}) = \min \left\{ 
\begin{aligned}
&c - 1 - x_2 + \exp\left(-\frac{x_1^2}{10}\right) + \left(\frac{x_1}{5}\right)^4,\\ 
&\quad
\frac{c^2}{2} - x_1x_2,
\end{aligned}
\right\}
\end{equation}
and the failure probability with  $c=3$ is $P_f \approx 3.48 \times 10^{-3}$. The changing thresholds $b^{(t)}$ of the adaptive subsets and the real and predicted performance functions over successive iterations are shown in Fig. \ref{fig_EXAMPLE1}. The sampling densities constructed of $N=3$ proposals are {depicted}. After {$T=12$} iterations, most of the generated sample points are observed to densely cover all over the {three} failure regions.

To investigate the accuracy of the $P_f$ estimation, the RRMSE of the failure probability estimates over 100 independent simulation runs is computed and presented in Table \ref{rrmse_values1}. In all the methods, we test different values of samples per iteration $K \in \{50, 100, 150, 200\}$, whereas $N \in \{2, 4, 6\}$. The initial
means are selected randomly $
\boldsymbol{\mu}^{(1)}_n=[\boldsymbol{\mu}_{n,1}^{(1)}, \boldsymbol{\mu}_{n,2}^{(1)},\dots \boldsymbol{\mu}_{n,d_x}^{(1)}] \text{ for } n = 1, 2, \ldots, N$.
Table \ref{Pf_values_cov} gives the best numerical results achieved by CE-PMC, SS-IS, and the presented method SAIS in terms of varying failure probabilities and the coefficient of variation $\delta_{\hat{P}_f}$. The variability of the estimates using both versions of SAIS (i.e., recycled and original) is very low compared to other methods, however, the decreasing trend of $\delta_{\hat{P}_f}$ with $\widehat{P}_f$ in the following examples is similar for all methods. Additionally, it is shown that CE-PMC and SS-IS produce biased $\widehat{P}_f$ estimates, and the bias tends to increase with decreasing number of random samples $K$ and decreasing failure probability. On the other hand, SAIS yields essentially unbiased estimates in most cases, since it is capable of populating the failure domains reasonably well.

\begin{table}[t]
    \centering
    \caption{S1: RRMSE values for SAIS algorithms and competitors with different sample sizes.}
    \label{rrmse_values1}
    \resizebox{\textwidth}{!}{%
    \begin{tabular}{@{}ccccccccccc@{}} 
        \toprule
        \toprule
        \multirow{2}{*}{Sample size} & \multicolumn{1}{c}{SS-IS} & \multicolumn{3}{c}{CE-PMC} & \multicolumn{3}{c}{SAIS} & \multicolumn{3}{c}{$\text{SAIS}_{\text{recycled}}$} \\ 
        \cmidrule(lr){2-2} \cmidrule(lr){3-5} \cmidrule(lr){6-8} \cmidrule(lr){9-11} 
        & \(N=1\) & \(N=2\) & \(N=4\) & \(N=6\) & \(N=2\) & \(N=4\) & \(N=6\) & \(N=2\) & \(N=4\) & \(N=6\)  \\ 
        \midrule
        $K=50$  & 0.836 & 1.220 & 1.140 & 0.970 & 0.282 & 0.155 & 0.098 & 0.263 & 0.154 & \textbf{0.085} \\
        $K=100$ & 0.808 & 0.922 & 0.724 & 0.581 & 0.251 & 0.088 & 0.066 & 0.236 & 0.085 & \textbf{0.063} \\
        $K=150$ & 0.778 & 0.818 & 0.646 & 0.469 & 0.242 & 0.071 & 0.055 & 0.232 & 0.070 & \textbf{0.054} \\
        $K=200$ & 0.733 & 0.768 & 0.522 & 0.372 & 0.214 & 0.037 & 0.033 & 0.209 & 0.037 & \textbf{0.029} \\
        \bottomrule
        \bottomrule
    \end{tabular}
    }
\end{table}

\vspace{-0.15cm}
\subsection{Example 2: Series system with four branches}

This renowned example from \cite{BAO2021107778,yang2020system} is selected to evaluate the computational details of the proposed SAIS method in problems problems with very small failure probability. Consider now a series system consisting
of four independent components with multiple failure modes. 
The performance function of this example can be read as

\begin{equation}
    S_2(\mbf{x}) = \min \left\{
\begin{aligned}
    &a + \frac{(x_1 - x_2)^2}{10} - \frac{x_1 + x_2}{\sqrt{2}}, \\
    &a + \frac{(x_1 - x_2)^2}{10} + \frac{x_1 + x_2}{\sqrt{2}}, \\
    &(x_1 - x_2) + \frac{b}{\sqrt{2}} + 1, \\
    &(x_2 - x_1) + \frac{b}{\sqrt{2}} + 1.
\end{aligned}
\right.
\end{equation}
The corresponding failure probability with $a=4$ and $b=7$ is $ P_f \approx 6.4 \times 10^{-5}$. The appeal of this example lies in its ability to highlight the discrepancy between the number of most probable points (MPPs) and the number of failure modes. Specifically, this example demonstrates a case where there are only two MPPs but four distinct failure domains. In this case, MPP-based
reliability methods can fail to capture the true complexity of the failure modes, as the count of MPPs does not accurately represent the underlying failure domains.

\begin{figure}[h!]
    \centering
    \includegraphics[width=0.4\linewidth]{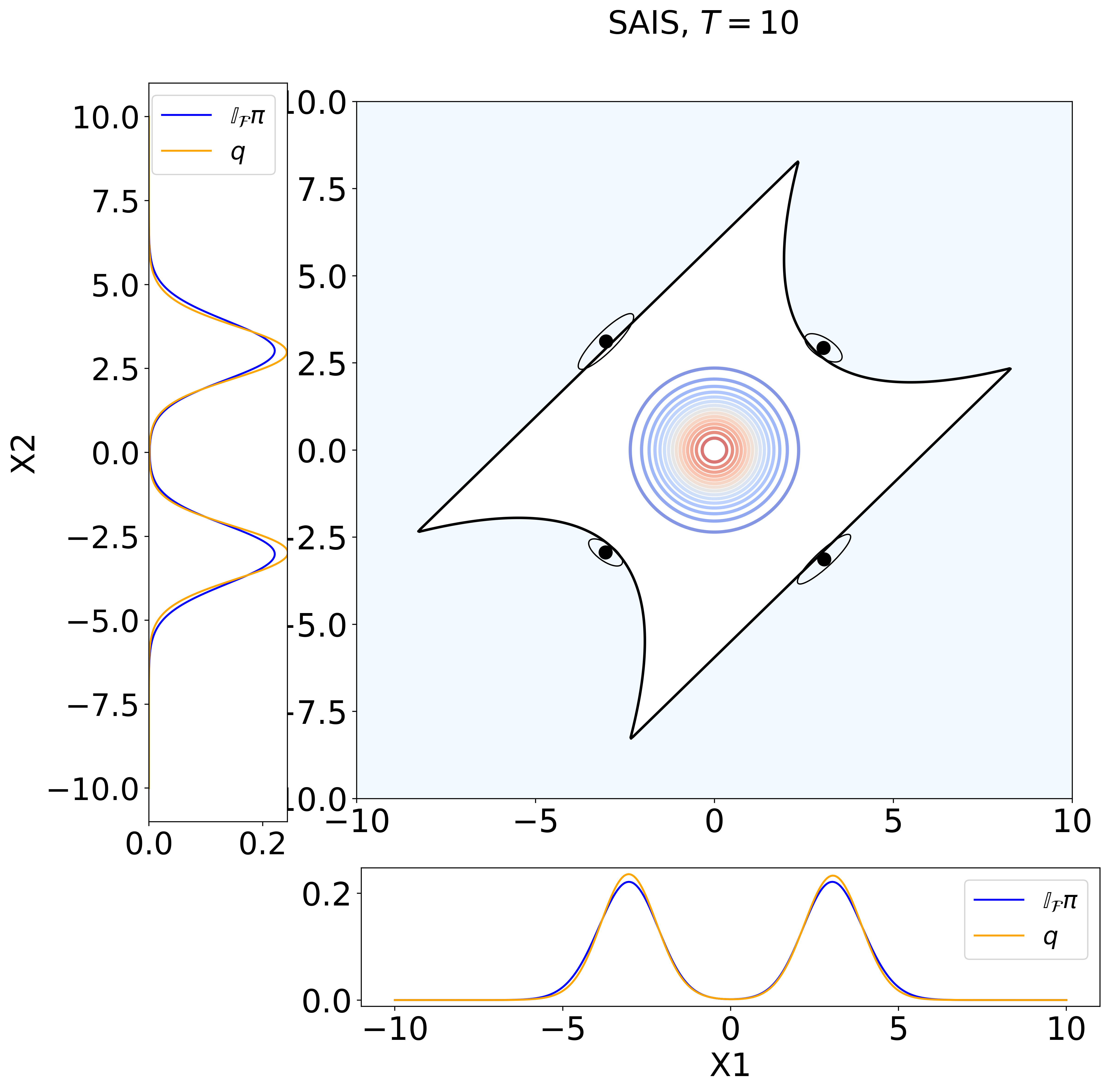}
    \includegraphics[width=0.4\linewidth]{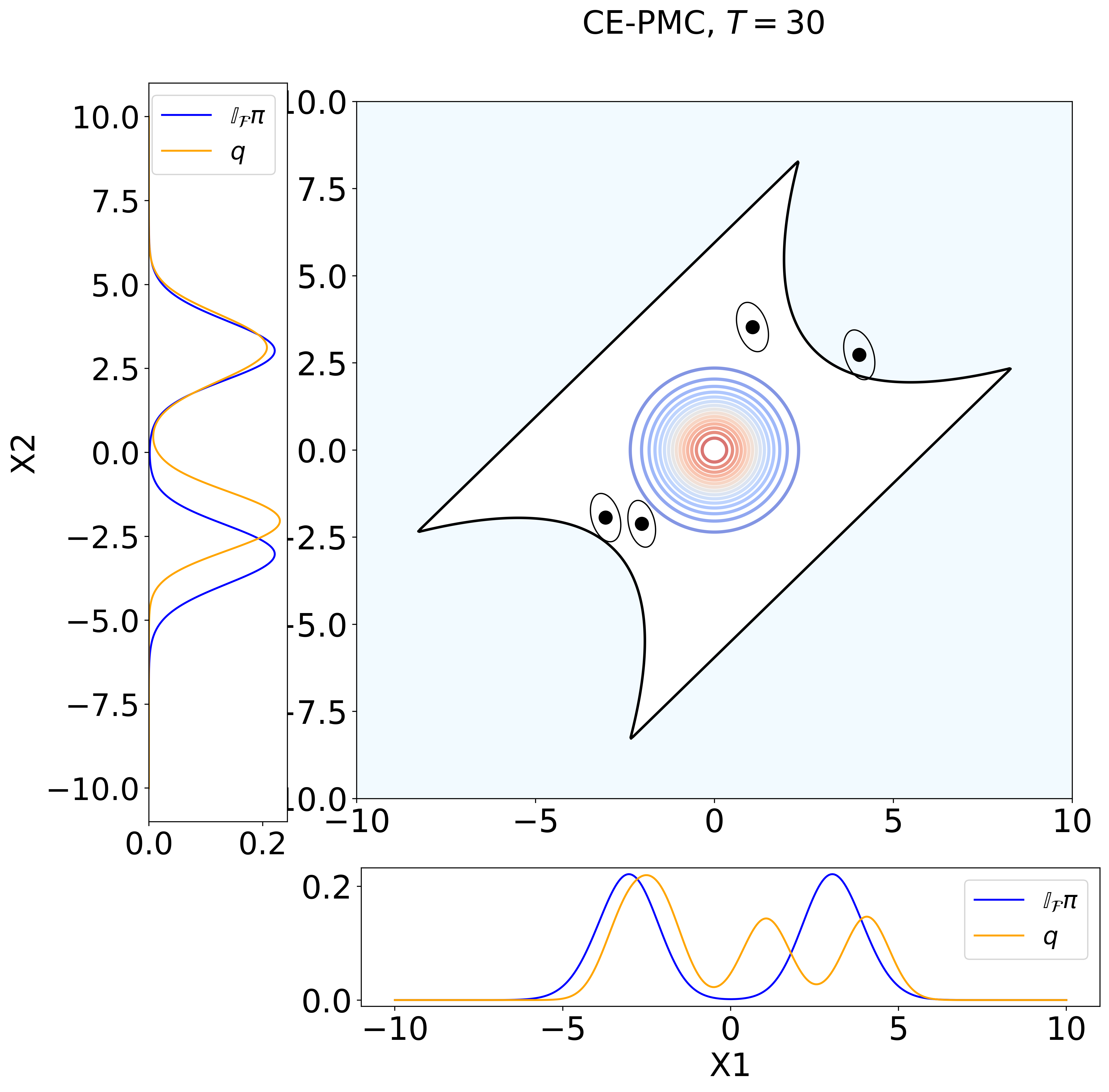}
    \caption{\textbf{Example 2.} The shaded region is the failure region $\mathcal{F}$ defined by the performance function $S_2(\mbf{x})$ (the black boundary). The central contour lines display the pdf $\pi(\mbf{x})$. The side plots present the marginal distributions of the target $\mathbb{I}_{\mathcal{F}}\pi(\mbf{x})$ (blue) and the proposal distributions $q(\mbf{x})$ (orange) along both axes.}
    \label{fig_subplots}
\end{figure}

Table \ref{rrmse_values2} shows the RRMSE results in the estimation of $P_f$. We can see that the proposed
scheme outperform all other methods for any value of $N$ and $K$. Moreover, we note that small values of $K$ lead to strong performance, whereas larger values of $K$ can be considered for improved performance without requiring more iterations $T$.  
When $N$ is chosen less than the number of failure modes (i.e., $N < 4$), the $N$ proposals adapt to the most probable points in SAIS. Fig. \ref{fig_subplots} shows the final means (black dots) and covariances (black ellipses) of $N=4$ proposals at the final iteration $T$ for SAIS and CE-PMC methods. It can be noted that the adaptation in CE-PMC was effective in recovering only two failure modes corresponding to the most probable points, while SAIS could successfully and automatically detect all the four failure modes.
Comparing SAIS with other methods, we see that both variants of SAIS give
smaller $\delta_{\hat{P}_f}$ at all target failure probabilities as shown in Table \ref{Pf_values_cov}. In addition, no further computation is required to determine the MPPs.
\begin{table}[h!]
    \centering
    \caption{S2: RRMSE values for SAIS algorithms and competitors with different sample sizes.}
    \label{rrmse_values2}
    \renewcommand{\arraystretch}{1.2}
    \resizebox{\textwidth}{!}{%
    \begin{tabular}{@{}ccccccccccc@{}} 
        \toprule
        \toprule
        \multirow{2}{*}{Sample size} & \multicolumn{1}{c}{SS-IS} & \multicolumn{3}{c}{CE-PMC} & \multicolumn{3}{c}{SAIS} & \multicolumn{3}{c}{$\text{SAIS}_{\text{recycled}}$} \\ 
        \cmidrule(lr){2-2} \cmidrule(lr){3-5} \cmidrule(lr){6-8} \cmidrule(lr){9-11} 
        & \(N=1\) & \(N=2\) & \(N=4\) & \(N=6\) & \(N=2\) & \(N=4\) & \(N=6\) & \(N=2\) & \(N=4\) & \(N=6\) \\ 
        \midrule
        $K=50$  & 0.523 & 6.580 & 5.267 & 4.394 & 0.458 & 0.159 & 0.128 & 0.451 & 0.157 & \textbf{0.116} \\
        $K=100$ & 0.298 & 4.900 & 5.164 & 4.207 & 0.417 & 0.148 & 0.062 & 0.402 & 0.144 & \textbf{0.057} \\
        $K=150$ & 0.235 & 4.351 & 4.968 & 2.990 & 0.371 & 0.070 & 0.064 & 0.369 & 0.065 & \textbf{0.063} \\
        $K=200$ & 0.212 & 3.018 & 4.528 & 1.803 & 0.336 & 0.045 & 0.035 & 0.335 & 0.044 & \textbf{0.033} \\
        \bottomrule
        \bottomrule
    \end{tabular}
    }
\end{table}

\subsection{Example 3: Modified Rastrigin problem}
 The third example is the modified Rastrigin function expressed as 
 \begin{equation}
   S_3(\mbf{x}) = 10 - \sum_{i=1}^{2} \left(x_i^2 - 5 \cos(2\pi x_i)\right).
\label{rastrigin}
\end{equation}The Rastrigin function is a classic example involving non-convex and non-connex
domains of failure (i.e., disjoint gaps of failure), commonly utilized as a benchmark problem to assess the performance of optimization algorithms, as seen in \cite{ZHANG2024103693,CHEN2022108639}. The difficulty arises from the highly complex failure domain, which is composed of several disconnected and scattered regions, coupled with a relatively high failure probability, $ P_f \approx 7.349 \times 10^{-2}$.

\begin{figure}[h!]
\centering
\includegraphics[width=0.6\textwidth]{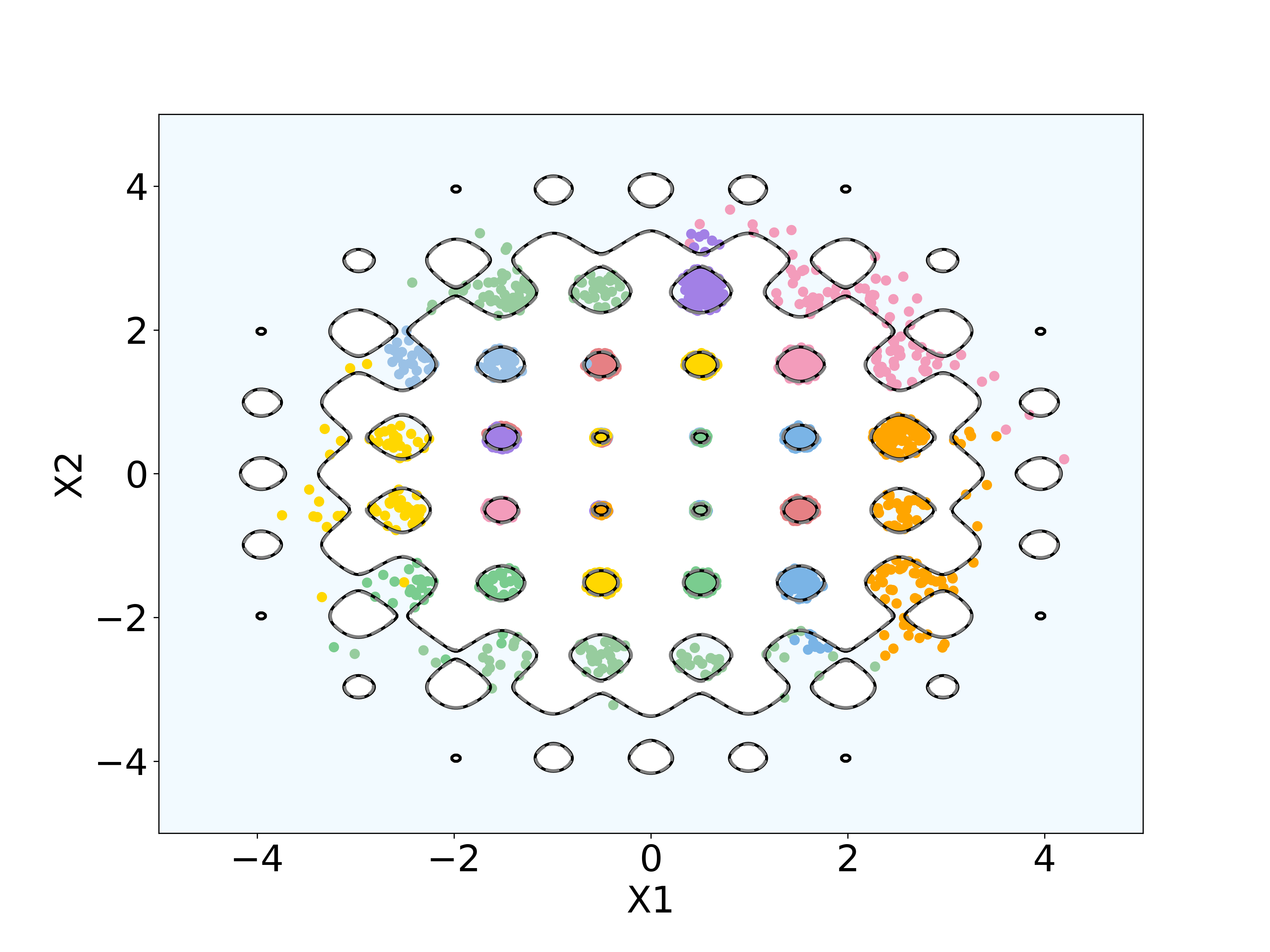}
\caption{\textbf{Example 3.} Evolution of $N=30$ proposal components over 10 iterations. The color code for each proposal samples is shown.}
\label{exp_rastigin}
\end{figure}

As observed again in Table \ref{Pf_values_cov}, the estimated failure probability by SS-IS ($\widehat{P}_f \approx 8.19\times 10^{-1}$) is quite diverged from the reference value  due to the generated samples being far away from the real $S_3$ curve, while a significant overlap between the reference value and SAIS estimation is evident.
CE-PMC becomes qualitatively similar to those of SAIS by calling more the performance functions with $T_{\text{CE-PMC}}=38$ iterations and adapting more proposals $N$ into the mixture model. The result of SAIS is very accurate, with RRMSE of only $0.034$ compared to the result obtained by SS-IS with RRMSE of $9.918$. This is particularly noticeable in Table \ref{rrmse_values3}.
In efficiency, both versions of SAIS require fewer $T_{\text{SAIS}}=5$ and call number than CE-PMC and SS-IS. 
The high accuracy demonstrates that the proposed algorithm can effectively explore and identify distinct failure regions, even in cases where the failure domain is highly noncontiguous, as seen in Fig. \ref{exp_rastigin}.
Therefore, the proposed SAIS shows remarkable advantages in terms of computing efficiency and estimation accuracy. 

\begin{table}[htbp]
    \centering
    \caption{S3: RRMSE values for SAIS algorithms and competitors with different sample sizes.}
    \label{rrmse_values3}
    \renewcommand{\arraystretch}{1.2}
    \resizebox{\textwidth}{!}{%
    \begin{tabular}{@{}ccccccccccc@{}} 
        \toprule
        \multirow{2}{*}{Sample size} & \multicolumn{1}{c}{SS-IS} & \multicolumn{3}{c}{CE-PMC} & \multicolumn{3}{c}{SAIS} & \multicolumn{3}{c}{$\text{SAIS}_{\text{recycled}}$} \\ 
        \cmidrule(lr){2-2} \cmidrule(lr){3-5} \cmidrule(lr){6-8} \cmidrule(lr){9-11} 
        & \(N=1\) & \(N=10\) & \(N=20\) & \(N=30\) & \(N=10\) & \(N=20\) & \(N=30\) & \(N=10\) & \(N=20\) & \(N=30\)  \\ 
        \midrule
        $K=50$  & 10.345 & 0.170 & 0.091 & \textbf{0.078} & 0.273 & 0.107 & 0.160 & 0.270 & 0.105 & 0.150 \\
        $K=100$ & 10.161 & 0.104 & 0.070 & 0.059 & 0.119 & 0.068 & 0.057 & 0.117 & 0.067 & \textbf{0.050} \\
        $K=150$ & 10.054 & 0.093 & 0.058 & 0.051 & 0.083 & 0.048 & 0.027 & 0.081 & 0.048 & \textbf{0.025} \\
        $K=200$ & 9.918 & 0.084 & 0.052 & 0.039 & 0.076 & 0.041 & 0.035 & 0.074 & 0.040 & \textbf{0.034} \\
        \bottomrule
        \bottomrule
    \end{tabular}
    }
\end{table}

\begin{table}[htbp]
    \centering
    \caption{Estimated failure probability and coefficient of variation for different methods across Examples 1--4.}
    \label{Pf_values_cov}
    \resizebox{\textwidth}{!}{%
    \begin{tabular}{@{}ccccccccc@{}} 
        \toprule
        \toprule
        \multirow{2}{*}{LSF} & \multicolumn{2}{c}{SS-IS} & \multicolumn{2}{c}{CE-PMC} & \multicolumn{2}{c}{SAIS} & \multicolumn{2}{c}{$\text{SAIS}_{\text{recycled}}$} \\ 
        \cmidrule(lr){2-3} \cmidrule(lr){4-5} \cmidrule(lr){6-7} \cmidrule(lr){8-9} 
        & $\widehat{P}_f$  & $\delta_{\widehat{P}_f}$ ($\%$)  & $\widehat{P}_f$  & $\delta_{\widehat{P}_f}$ ($\%$)  & $\widehat{P}_f$  & $\delta_{\widehat{P}_f}$ ($\%$)  & $\widehat{P}_f$  & $\delta_{\widehat{P}_f}$ ($\%$)   \\ 
        \midrule
    
        $S_1$  & $5.80\times 10^{-3}$ &12.35  & $3.84\times 10^{-3}$ &30.00  & $3.46\times 10^{-3}$ &1.88  & $3.47\times 10^{-3}$ &1.76 \\
        $S_2$  & $6.35\times 10^{-5}$ &21.76  & $5.26\times 10^{-5}$ &44.95  & $6.30\times 10^{-5}$ &4.70  & $6.27\times 10^{-5}$ &4.62 \\
        $S_3$  & $8.19\times 10^{-1}$ &5.94  & $7.29\times 10^{-2}$ &5.28  & $7.36\times 10^{-2}$ &1.73  & $7.35\times 10^{-2}$ &1.60 \\
        $S_4 (d_x=20)$ & $5.72\times 10^{-4}$ &13.91  & $2.26\times 10^{-4}$ &46.40  & $2.30\times 10^{-4}$ &2.00  & $2.33\times 10^{-4}$ &1.03 \\
        $S_4 (d_x=40)$ & $1.10\times 10^{-4}$ &23.96  & $2.41\times 10^{-4}$ &50.22  & $2.32\times 10^{-4}$ &2.90  & $2.28\times 10^{-4}$ &2.54 \\
        $S_4 (d_x=60)$ & $2.70\times 10^{-6}$ &30.71  & $3.65\times 10^{-4}$ &54.20  & $2.29\times 10^{-4}$ &4.60  & $2.31\times 10^{-4}$ &4.21 \\
        $S_4 (d_x=80)$ & $5.56\times 10^{-12}$ &50.34  & $4.83\times 10^{-4}$ &61.70  & $2.13\times 10^{-4}$ &6.50  & $2.52\times 10^{-4}$ &5.15 \\
        \bottomrule
        \bottomrule
    \end{tabular}
    }
\end{table}

\subsection{Example 4: Numerical comparison in variable dimension}
The final example aims to demonstrate the ability of SAIS to handle high dimensional problems. We compare CE-PMC
and SS-IS with SAIS, as in the previous examples. The performance function is expressed as a linear function of independent standard normal variables and is given by
\begin{equation}
\label{EX4}
    S_4(\mbf{x}) = \gamma - \frac{1}{\sqrt{d_x}} \sum_{i=1}^{d_x} x_i.
\end{equation}
The failure probability for the performance function (\ref{EX4}) is $P_f = \Phi(-\gamma)$ independent of the dimension $d_x$,
where $\Phi$  is the cumulative distribution function (CDF) of the standard normal distribution. Here, we choose \( \gamma = 3.5 \), corresponding to a rare-event probability of  $\Phi(-3.5) \approx 2.33 \times 10^{-4}$, and consider various dimensionalities $d_x \in \{5, 10, 20, 30, 40, 50, 60, 70, 80, 90, 100\}$. 
For fair comparison, we run SAIS and CE-PMC with $N = 5$ and $K = 3000$ keeping fixed the initial means $\boldsymbol{\mu}_n^{(1)}\in [-1, 1]^{d_x} \quad \text{for } n = 1, 2, \ldots, N$, while the initial covariances are set isotropic, \( \mbf{\Sigma}_n^{(1)} = \sigma^2 \mbf{I}_{d_x} \), where \( \sigma = 1 \). In this example, the particular choice of the quantile parameter $\rho=0.2$ seems to be a good
choice for all dimensions considered.

\begin{figure*}[t]
    \centering

    \includegraphics[width=0.45\linewidth]{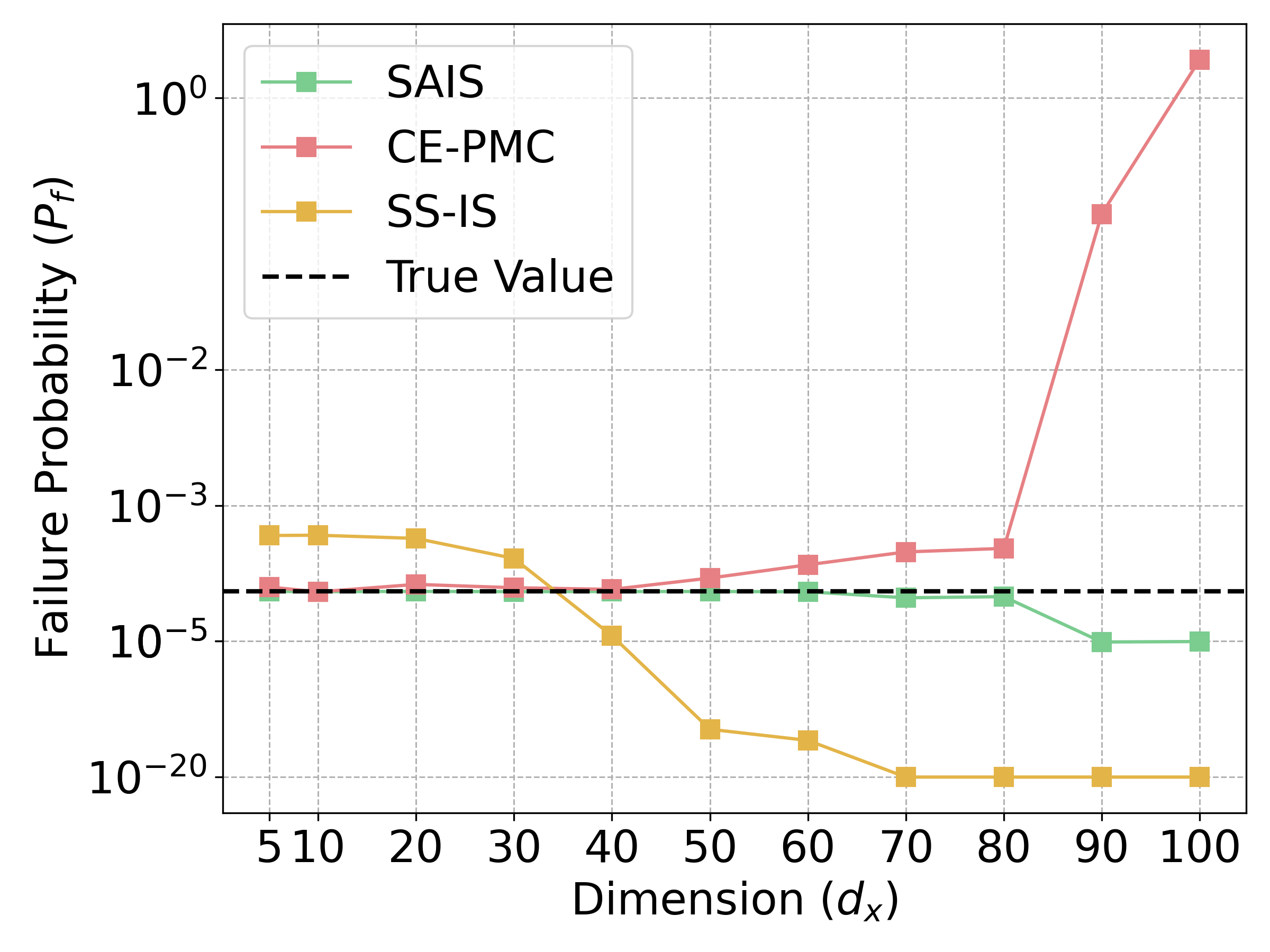}
    \includegraphics[width=0.45\linewidth]{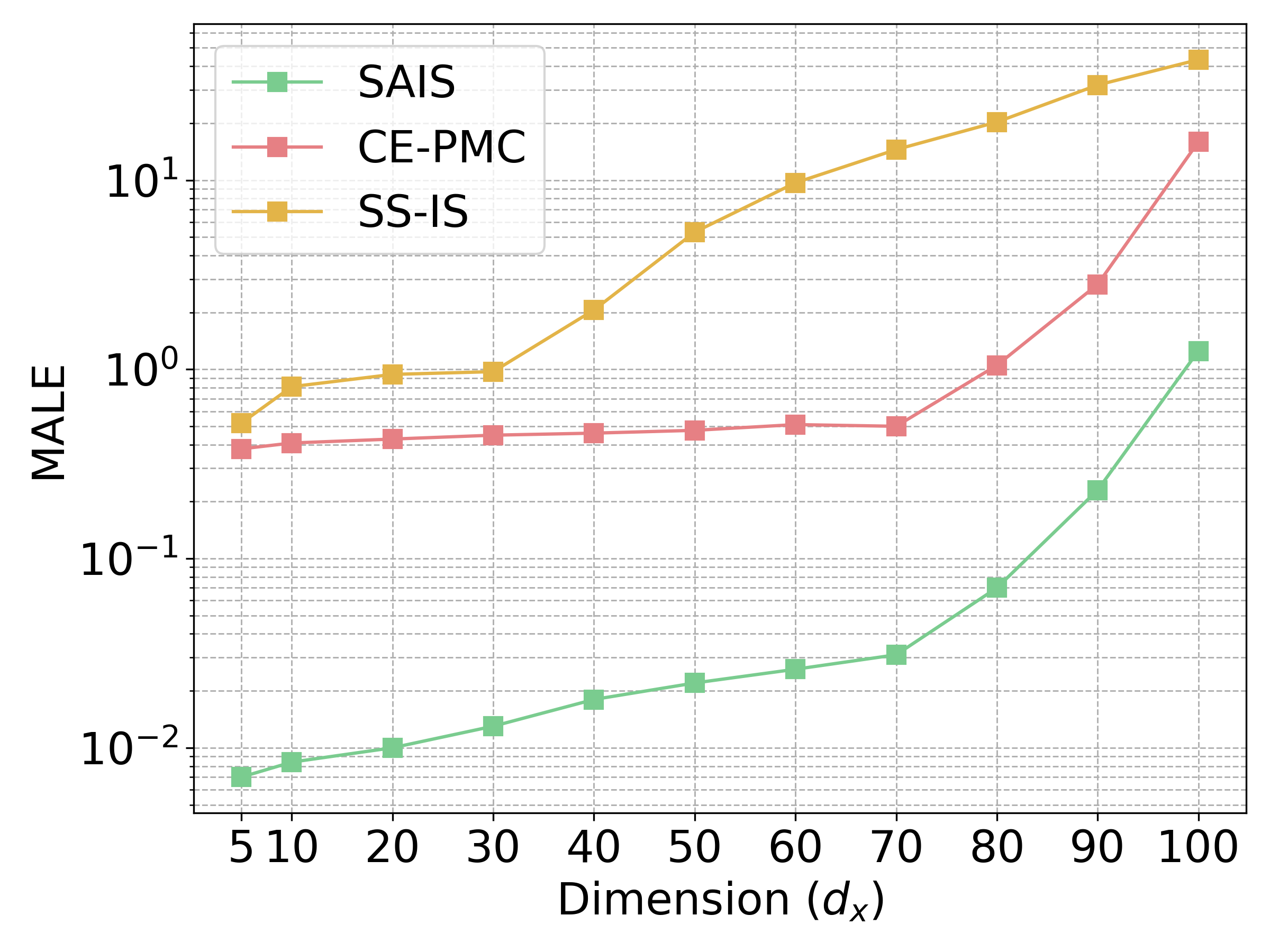}
    
    \caption{\textbf{Example 4.} Comparison of rare event estimation methods across increasing dimensions. The left plot shows the estimated failure probability $P_f$ as a function of dimensionality $d_x$ for three methods. The true failure probability (black dashed line) = $2.33 \times 10^{-4}$. The right plot presents the mean absolute log error (MALE) as a function of $d_x$.}
    \label{multi_EX}
\end{figure*}

The estimation results are listed in Table \ref{Pf_values_cov}. Fig. \ref{multi_EX} demonstrates the results of utilizing the three methods to estimate the failure probability in high dimensions. The left panel of Fig. \ref{multi_EX} plots the estimated $P_f$ versus dimensions $d_x$. The dashed line corresponds to the true reference value based on $10^{10}$ MC samples. SAIS exhibits the most stable performance and robsutness to dimensionality, while SS-IS method severely underestimates $P_f$ for high dimensions and fails to capture the rare event probability. 
The SAIS robustness can be attributed to its covariance shrinkage approach and the regularized empirical covariance estimation described in Eq. (\ref{Shrink_cov}).
Note that for $d_x \in [5,50]$ both SAIS and CE-PMC produce accurate estimates. For higher values $d_x > 50$, however, CE-PMC starts to degenerate and diverge sharply, resulting in a pronounced overestimation in high dimensions. 
The right panel of Fig. \ref{multi_EX} represents the relative mean absolute logarithmic error (MALE) \cite{Tofallis01082015} of all three methods as a function of the average number of $R$ runs
\[
\text{MALE} = \frac{1}{R} \sum_{i=1}^{R} \left| \log \left( \frac{\hat{P}_f^{(i)}}{P_f} \right) \right|,
\]
where the subscript $i$ here denotes the run or the number of evaluations corresponding to the estimates.
The SAIS algorithm clearly and consistently outperforms both CE-PMC and SS-IS: the relative MALE increases at a rate much slower than SS-IS, in particular.
In CE-PMC, the MALE remains nearly constant for dimensions $d_x \leq 50$, at the expense of high computational overhead, i.e., converges in $T_{\text{CE-PMC}} \geq 100$ iterations. This suggests that CE-PMC maintains stable performance in moderate dimensions at high cost but suffers from deteriorating accuracy in higher dimensions due to weight degeneracy and slow adaptation.
In contrast, SAIS exhibits a slow and controlled increase of MALE over the entire range of $d_x$, maintaining significantly lower error values compared to CE-PMC and SS-IS. This trend highlights the ability o SAIS to mitigate high-dimensional bias-variance tradeoffs effectively with only $T_{\text{SAIS}} \in [7, 9]$ iterations to meet stopping criterion. 
In this example, and all previous ones, $T=10$ iterations are sufficient, which reserves most of the computational effort toward increasing $N$ or $K$. This is particularly critical during the first adaptation iteration, where the initial sampling densities are still poor, and in the final iterations, where greater accuracy of the failure approximation is essential. 
Another observation is that despite the fact SAIS converges quickly to failure probability, it takes much longer to shrink the estimator around failure domains in higher dimensions.

\subsection{Discussion on the numerical results}
In light of the above examples, the following observations can be made. First, SAIS demonstrates superior accuracy across all benchmarks, competing favorably with existing reliability analysis methods such as SS-IS and CE-PMC. It achieves better results in 91.67\% of tested cases, and its average performance surpasses even the best-case results of the reference methods. This advantage extends across both low- and high-dimensional problems and encompasses varying degrees of nonlinearity and complexity. In addition to accuracy, SAIS exhibits robustness to problem complexity, as evidenced by the RRMSE values across all four experiments, which show relatively uniform performance regardless of the underlying limit state function. This consistency suggests that SAIS is largely insensitive to the shape or complexity of the target failure domain, a critical advantage in reliability analysis where problem characteristics can vary significantly. Furthermore, SAIS  reduces the number of expensive performance function evaluations and model
simulations due to its multilevel and proposal adaptation strategies. Unlike conventional methods, it requires only a small, informative subset of samples that efficiently target regions of high failure likelihood, resulting in predicted values that closely approximate the true value across problem dimensions.

\section{Conclusion}
\label{conc}
The estimation of rare event or failure probability in high-dimensions is of significant interest across many areas such as reliability analysis, risk assessment, and safety engineering. The proposed SAIS algorithm provides a flexible and robust framework  for adapting a mixture of importance distributions, allowing highly accurate approximation of rare event probabilities. 
The update mechanism facilitates early stabilization of the mixture parameters. Therefore, it requires few iterations with relatively small sample sizes at each iteration. In the numerical examples, we have shown that the SAIS algorithm permits estimating failure probabilities accurately from $10^{-2}$ to $10^{-5}$ with relatively low RRMSE and MALE results. SAIS becomes more efficient compared with SS-IS and CE-PMC as the target rare event probability gets smaller. Moreover, the gain in performance by SAIS is more pronounced in larger dimensions and complex curvatures of performance functions. In summary, extending the capabilities of Monte Carlo simulation with our proposed procedures provides an avenue for analyzing larger, more complex, and nonlinear systems in the context of low-probability events and their reliability assessment.

\section*{Acknowledgements}
The authors gratefully acknowledge the support from the EPSRC Centre for Doctoral Training in Mathematical Modeling, Analysis and Computation (MAC-MIGS) funded by the UK Engineering and Physical Sciences Research Council (grant EP/S023291/1) and ARL/ARO under grant W911NF-22-1-0235.

\end{document}